\documentclass[ amsmath,amssymb,nofootinbib
]{revtex4-1}

\usepackage[table]{xcolor}
\usepackage{graphicx}

\usepackage{bm}
\usepackage{subcaption}
\usepackage{comment}
\usepackage{extarrows}
\usepackage{dcolumn}
\usepackage{mathrsfs}

\usepackage{rotating}

\usepackage{ulem}

\def\bal#1\eal{\begin{align}#1\end{align}}
\def\mbf#1{\mbox{\boldmath $#1$}}

\newtheorem{theorem}{Theorem}

\newcommand{\be}{\begin{equation}}
\newcommand{\ee}{\end{equation}}
\newcommand{\bea}{\begin{eqnarray}}
\newcommand{\eea}{\end{eqnarray}}
\newcommand{\beaa}{\begin{eqnarray*}}
\newcommand{\eeaa}{\end{eqnarray*}}

\newcommand{\bN}{\mathbb{N}}
\newcommand{\bR}{\mathbb{R}}

\newcommand{\bQ}{{\bf Q}}

\newcommand{\bS}{{\bf S}}

\newcommand{\bw}{{\bf w}}

\newcommand{\bx}{{\mbf x}}

\newcommand{\bc}{{\bf c}}
\newcommand{\bfe}{{\mbf f}}

\newcommand{\br}{{\mbf r}}
\newcommand{\tM}{{\tilde M}}
\newcommand{\tnu}{{\tilde \nu}}
\newcommand{\tJ}{{\tilde J}}

\newcommand{\p}{\partial}
\newcommand{\non}{\nonumber}

\newcommand{\dk}{{\rm \dim ker}\, }
\newcommand{\detA}{{\rm det}\,  {\bf A}}
\newcommand{\gs}{{\Gamma_s}}
\newcommand{\gbs}{{\bar{\Gamma}_s}}

\newcommand{\red}[1]{\textcolor{black}{#1}}

\newcommand{\corr}[1]{\textcolor{black}{#1}}

\newcommand{\green}[1]{\textcolor{black}{#1}}

\definecolor{mygreen}{rgb}{0.55, 0.71, 0.0}
\newcommand{\TO}[1]{\textcolor{black}{#1}}

\begin{document}

\title{Structural Bifurcation Analysis in Chemical Reaction Networks}

\author{Takashi Okada$^{1}$, Je-Chiang Tsai$^{2}$, and Atsushi Mochizuki$^{1,3}$}

\affiliation{
$^1$Theoretical Biology Laboratory, RIKEN, Wako 351-0198, Japan \\
$^2$ Department of Mathematics, National Tsing Hua University,
 Hsinchu 300, Taiwan \\
$^3$CREST, JST
4-1-8 Honcho, Kawaguchi 332-0012, Japan
}
 \email{E-mail address: takashi.okada@riken.jp}
\begin{abstract}


In living cells, chemical reactions form a complex network. Complicated dynamics arising from such networks are the origins of biological functions. We propose a novel mathematical method to analyze bifurcation behaviors of a reaction system from the network structure alone. The whole network is decomposed into subnetworks based on ``buffering structures". For each subnetwork, the bifurcation condition is studied independently, and the parameters that can induce bifurcations and the chemicals that can exhibit bifurcations are determined. We demonstrate our theory using hypothetical and real networks.

\end{abstract}

\maketitle



\section{ Introduction}
\begin{figure}[h!tb]
\includegraphics[width=8.5cm,clip]{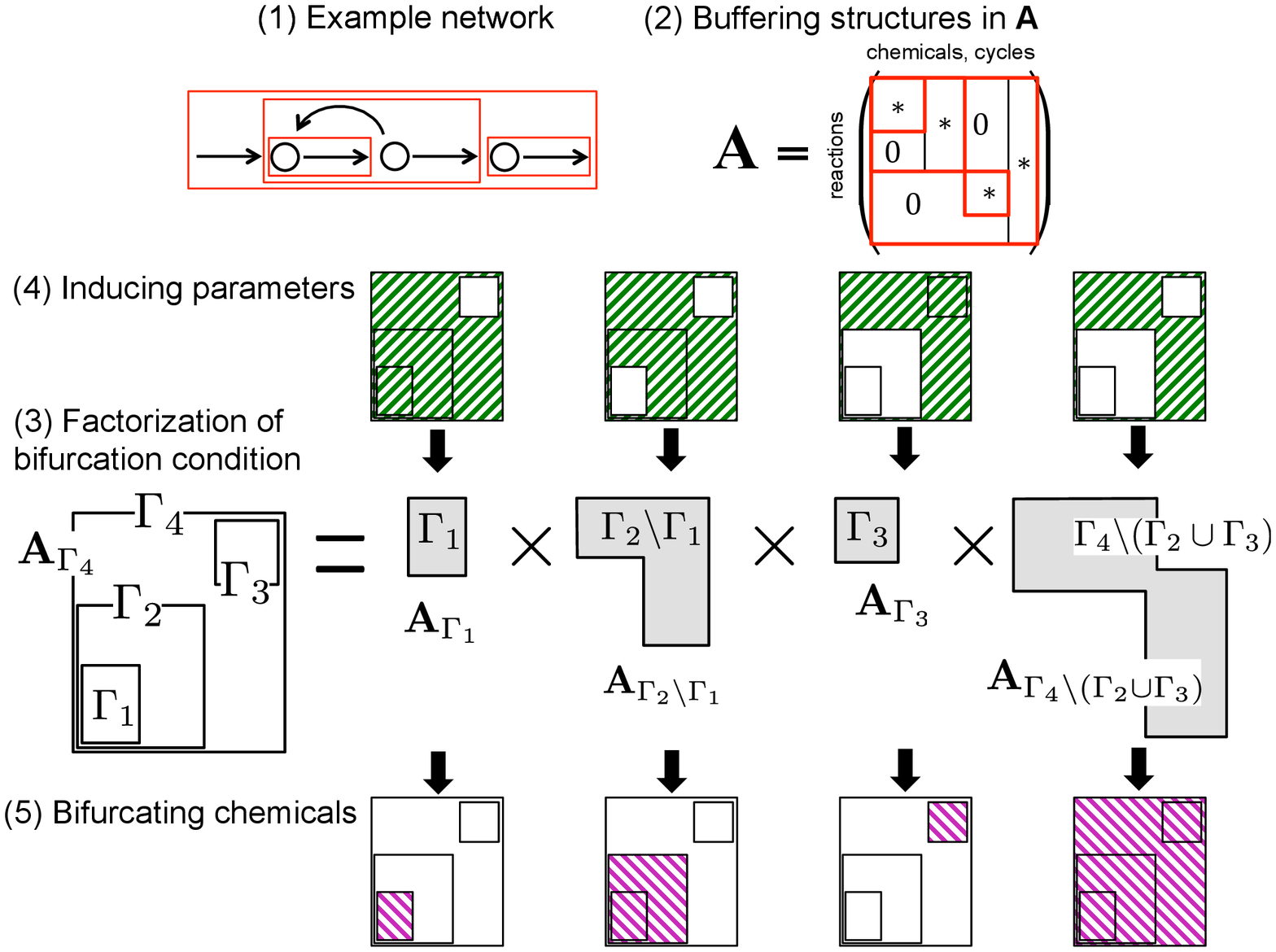}
\caption{Summary of the structural bifurcation analysis. 
(1)  { Buffering structures (red boxes) in an example network.} 
(2) { Buffering structure corresponds to nonzero square blocks  in {\bf A}}. 
\corr{
(3) Bifurcations  {in the whole system are} governed by a product of buffering structures with subtraction of their inner buffering structures. 
(4) For each subnetwork, { parameters in shadowed area can induce bifurcations associated with the subnetwork}.
(5) For each subnetwork, { chemicals in shadowed area exhibit bifurcations}. 
}
}
\label{fig:ov}
\end{figure}

In a living cell, a large set of chemical reactions are connected by sharing their products and substrates, and constructing a large network.
Biological functions are believed to  arise from dynamics of chemical concentrations based on the networks. 
It is also considered that regulations and adaptations of biological systems are realized by modulations of amount/activities of enzymes mediating reactions.
In previous studies \cite{Mochizuki, OM}, we developed a mathematical method, by which the sensitivity responses of chemical reaction networks to perturbations of enzyme amount/activities are determined from network structures alone. 
Our method is based on an augmented matrix {\bf A} (see Eq. \eqref{e:A_temp30}), in which the distribution of nonzero entries directly reflects  network structures. 
One of the striking result is the {\it law of localization} \cite{OM}. \corr{ A substructure (subset of chemicals and reactions) in a reaction network satisfying a topological condition is called {\it buffering structure} (Fig.~\ref{fig:ov}(1), (2)), and  has the property that  perturbations of reaction rate parameters inside a buffering structure influence   only (steady-state) concentrations and fluxes inside this structure, and the outside remains unchanged under the perturbations.}

\corr{In this paper, we study another aspect  governed by buffering structures: bifurcation behaviors \red{of}  reaction systems.}
We prove that  the determinant of the Jacobian matrix {\bf J} of  a reaction system is equivalent to that of the augmented matrix {\bf A}  for the corresponding network structure. 
Based on this equivalence, we study steady-state bifurcations of  reaction systems from network structures. In this paper, our usage of  ``parameter'' always means a parameter associated with a reaction rate. 

From the {\it  structural bifurcation analysis} based on the matrix {\bf A}, we  obtain the following general results on steady-state bifurcations in reaction networks. 

\corr{
(i) {\it Factorization}:  {\bf A} is factorized into submatrices based on the buffering structures (Fig.~\ref{fig:ov}(3)).
It implies that { bifurcation behaviors in} a complex network can be studied  by decomposing it into smaller subnetworks, which are buffering structures with subtraction of their inner buffering structures. 
For each subnetwork, the condition of bifurcation occurrence is determined from \red{the} structure of {the subnetwork}. 
}

\corr{
(ii) {\it Inducing parameters}: 
For each subnetwork, bifurcation is induced by parameter changes which are neither in buffering structures inside the subnetwork nor those non-intersecting with the subnetwork (Fig.~\ref{fig:ov} (4)).
}

\corr{
(iii) {\it Bifurcating chemicals {(and fluxes)}}:  
When the condition of bifurcation associated with a subnetwork is satisfied, the bifurcation of steady-state concentrations {(and fluxes)} \red{appears} only inside the (minimal) buffering structure containing the subnetwork (Fig.~\ref{fig:ov} (5)).
}


These findings make it possible to study behaviors of a whole reaction system including multiple bifurcations based on inclusion relations of buffering structures.
We apply our method to {hypothetical and real} networks, and demonstrate the practical usefulness to analyze behaviors of complex systems.


\section{Structural sensitivity analysis.---}We label chemicals by  $m \,(m=1,\cdots,M)$ and reactions by  $n\,(n=1,\cdots,N)$.
A state of a spatially homogeneous chemical reaction system is specified by the concentrations $x_m(t)$,
and obeys the differential equations \cite{Mochizuki,F1,F2}:
\begin{align} \label{e:ode_app}
\frac{d x_m}{dt}= \sum_{n=1}^N \nu_{mn} r_n(\bx;k_n),   \; m=1,\ldots,M.
\end{align}
Here, the $M\times N$ matrix $\nu$ is called the stoichiometric matrix,
and its component $\nu_{mn}$ is defined as follows:
Let the stoichiometry of the $n$-th reaction  among  chemical molecules $X_m$ be given by
$
   \sum_{m=1}^M y^n_m X_m \rightarrow \sum_{m=1}^M \bar{y}^n_m X_m 
$
Then the  $\nu_{mn}$ is given by
$
\nu_{mn}=  \bar{y}^n_m -y^n_m.
$
The reaction rate function (flux) $r_n$ depends on the concentration vector $\bx$ and
also on rate parameters $k_n$.

To present the key idea, we start by assuming that \corr{ the stoichiometric matrix  $\nu$ does not have nonzero cokernel vectors,} which in turn implies that \corr{${\rm rank}(\nu) = M \le N$} and the steady state concentration $\bx^*$ and fluxes $\br^*$
are continuous functions of rate parameters $\{ k_n \}_{n=1}^N$.  
The general case will be presented in the Supplemental Material (SM).
For  steady state $\bx^*$,
one can choose $\{\mu^\alpha\}_{\alpha=1}^{N-M}$  such that
the corresponding steady state flux $\br^*$ \corr{is} expressed,
in terms of the basis $\{{\mbf c}_\alpha\}_{\alpha=1}^{N-M}$ of the kernel space  of $\nu$, as
\begin{align} \label{ker_temp30}
   \br^* = \sum_{\alpha=1}^K \mu^\alpha {\mbf c}_\alpha, \quad K := N-M = \dk  \nu.
\end{align}
%

Now we review the {\it structural sensitivity analysis} \cite{Mochizuki} and  the {\it law of localization} \cite{OM}. 
Under the  perturbation $k_{\hat{n}} \to k_{\hat{n}} + \delta k_{\hat{n}}$ ($\hat{n} = 1,\ldots,N$),
the corresponding concentration changes $\delta_{\hat{n}} x_m$ and the flux changes $\delta_{\hat{n}} r_n$
at the steady state $\bx^*$ are  determined simultaneously by solving the following equation
\begin{align} 
{\bf A} \left(
\begin{array}{cccc}
\delta_1 {  \mbf x} &  \ldots  &\delta_{N} { \mbf  x}  \\\hline
\delta_{1} {  \mbf \mu} &  \ldots & \delta_{N} {  \mbf \mu}  \\
\end{array}
\right)
= - {\rm diag}\Big(\frac{\partial r_1}{\partial k_1}\delta k_1,  \  \ldots,  \  \frac{\partial r_N}{\partial k_N} \delta k_N \Big),\nonumber
\end{align}
where
the horizontal line denotes the structure of block matrices,
the derivatives are evaluated at the steady state $\bx^*$,
$\delta_{\hat{n}}{\mbf \mu} = (\delta_{\hat{n}}\mu^1,\ldots,\delta_{\hat{n}}\mu^K)^T$
is the change of the coefficients $\mu^\alpha$ in (\ref{ker_temp30})
under the perturbation: $k_{\hat{n}} \to k_{\hat{n}} + \delta k_{\hat{n}}$,
and the matrix ${\bf A}$ is given by
\begin{align} \label{e:A_temp30}
   {\bf A} = \Big( \frac{\p \br}{\p \bx}\Big|_{\bx=\bx^*} -{\mbf c}_1 \ldots -{\mbf c}_K \Big).
\end{align}
Here the entry $\partial r_n/\partial x_m$ in the left part of $\bf A$ is given by 
\begin{align}
\begin{cases}
\frac{\partial  {\color{black}r}_n}{\partial x_m}     \neq 0 \  \    {\rm  if}\  x_m \ {\rm influences \ reaction} \ n,  \\
  \frac{\partial {\color{black}r}_n}{\partial x_m} =0\  \  {\rm otherwise.}\ 
  \end{cases} 
\label{reg}
\end{align}
Note that whether each entry of {\bf A} is zero or nonzero is determined structurally, that is,  independently of  detailed forms of rate functions and quantitative values of concentrations and parameters.

For a given network $\Gamma$, we consider a pair $\Gamma_s=({\mathfrak m}, {\mathfrak n})$  of a chemical subset ${\mathfrak m}$ and a reaction subset ${\mathfrak n}$ satisfying the condition that ${\mathfrak n}$ includes all reactions influenced by {chemicals} in ${\mathfrak m}$ {(in the sense of \eqref{reg})}. 
We  call  such a $\Gamma_s=({\mathfrak m}, {\mathfrak n})$ as {\it output-complete}. 
For a subnetwork $\Gamma_s = (\mathfrak m, \mathfrak n)$, we define the {\it index} $\chi(\Gamma_s)$ as
\bal
\chi(\Gamma_s) = |\mathfrak m| -| \mathfrak n| + (\# cycle).
\eal
Here, $|\mathfrak m|$ and  $| \mathfrak n|$ are the numbers of elements of the sets $\mathfrak m$ and  $\mathfrak n$, respectively.
The $\# cycle$ is the number of independent  kernel vectors of the matrix $\nu$ whose supports are contained in $\mathfrak n$.
In general, $\chi(\Gamma_s)$ is non-positive (see \cite{OM}).
Then a {\it buffering structure} is defined as an output-complete subnetwork $\Gamma_s$ with $\chi(\Gamma_s) =0$.

Suppose that $\Gamma_s=({\mathfrak m},{ \mathfrak n})$ is a buffering structure.
By 
 permutating the column and row indices, the matrix  $\bf A$ can be written as  \red{follows:}~\cite{OM}
\bal
{\bf A}= \begin{array}{c }
\ _{ |{\mathfrak n}|}\big{\updownarrow} \\
\\ 
\end{array}
\overset{\overset{ |{\mathfrak m}|+ (\# cycle) }{\xlongleftrightarrow[]{\hspace{3em}}}\quad \quad \quad \quad }{\left(
\begin{array}{ccc|c }
&\underset{square}{\mbox{\smash{ ${\bf A}_{\Gamma_s}$}}}&&\ \ \ \ \mbox{\smash{\Large $*$}}\ \  \   \\ \hline 
&\mbox{\smash{ $\bf 0$}}&& \ \ \ \mbox{\smash{ ${\bf A}_{\bar \Gamma_s}$}}\ \ \  
\end{array}
\right)},\label{Agamma} 
\eal
where the rows (columns) of ${\bf A}_{\Gamma_s}$ are associated with the reactions (chemicals and cycles) in $\Gamma_s$. Similarly, those of ${\bf A}_{\bar{\Gamma}_s}$ are associated with the complement  ${\bar\Gamma}_s :=\Gamma\backslash \Gamma_s$.

{ The  {\it law of localization } \cite{OM} is a direct consequence of the block structure \eqref{Agamma} of the matrix {\bf A}: } {\it {Steady-state  concentrations and  fluxes} outside of a buffering structure does not change under any rate parameter perturbations in $\mathfrak n$.}  In other words, all effects of perturbations of $k_{\hat{n}}$ in $\mathfrak n$ are indeed localized within $\Gamma_s$.


\section{ Structural bifurcation analysis.---}
We shortly sketch the conventional bifurcation analysis. Set $J:=\nu \frac{\p \br}{\p \bx}|_{\bx=\bx^*}$. 
Let the eigenvalues $\{\sigma_m\}_{m=1}^M$ of $J$ be ordered so that
${\rm Re}\, \sigma_1 \ge {\rm Re}\, \sigma_2 \ldots$.
Then the state $\bx^*$ is stable if all ${\rm Re}\, \sigma_m < 0$,
whereas, the state $\bx^*$ is unstable if ${\rm Re}\, \sigma_1 > 0$.
Moreover,
a bifurcation occurs if ${\rm Re}\, \sigma_1$ changes its sign
as some parameter $k_{\hat n}$ is varied through some critical value, say $\bar{k}_{\hat n}$.
Since $\det J  = \prod_{m=1}^M\sigma_m$,
a bifurcation  occurs if $\det J$ changes its sign
as $k_{\hat n}$ is varied through $\bar{k}_{\hat n}$.
Thus, the study of  the onset of bifurcation is reduced to the search of the zeros of $\det J$.

Now, we explain {\it structural bifurcation analysis}. 
The following relation is the key to our method: 
\begin{align} \label{e:e:main_20_temp10}
    \det J  \propto \det {\bf A},
\end{align}
where 
the proportionality constant depends only on the stoichiometric matrix $\nu$. See the SM for the proof.
Then (\ref{e:e:main_20_temp10}) implies that 
the study of  the onset of bifurcation of system (\ref{e:ode_app})
can be reduced to the search of the zeros of $\detA$.
Further, 
the existence of the buffering structure $\Gamma_s$ of the network $\Gamma$
guarantees the following relation
\corr{
\begin{equation} \label{e:e:main_20_temp50}
     \det J   \propto \det {\bf A}_{\Gamma_s} \cdot \det {\bf A}_{\bar \Gamma_s}.
\end{equation}
}
An extension of (\ref{e:e:main_20_temp50}) to a nested sequence of buffering structures 
$\Gamma_1\subset \ldots \subset \Gamma_L$ gives the \red{relation}
$\det J \propto \det {\bf A}_{\Gamma_1} \prod_{s=1}^{L-1}  \det {\bf A}_{{\Gamma_{s+1}} \backslash {\Gamma_s}}$.

\corr{
There are several implications of the factorization (\ref{e:e:main_20_temp50}). 
First, from the equality $\det J=0$, \red{we have either $\detA_{\Gamma_s}=0$ or  $\det {\bf A}_{\bar \Gamma_s}=0$.}  
In other words, the possibility of bifurcation occurrences in \red{the} whole system can be studied by examining the possibility for each of the subnetworks from their structures. } 

\corr{
Second, from the law of localization \red{(see the SM for details)}, $\detA_{\bar \Gamma_s}$ depends only on parameters outside $\Gamma_s$, whereas  $\detA_{\Gamma_s}$ depends on parameters in the whole network $\Gamma$. 
Thus, bifurcations associated with $\detA_\gbs$ are triggered only by tuning parameters in $\bar\Gamma_s$, while bifurcations associated with $\detA_\gs$ \red{can be}  induced by both parameters in $\gs$ and those in $\gbs$ \red{(see inducing parameters in Fig.~1 (4))}. In particular, in the former case, critical values (values at bifurcation points) of  parameters in $\gbs$ are independent of parameters in $\Gamma_s$. 
}

\corr{
Third,  as shown in the SM,  the null vector of the Jacobian $J$ at  a bifurcation point with $\detA_\gs=0$ has  nonzero components only for  chemicals in $\Gamma_s$. This implies that only chemicals in $\Gamma_s$  exhibit bifurcations at this bifurcation point. By contrast,  
\red{for} bifurcations associated with $\detA_\gbs$,  all chemicals in $\Gamma$  exhibit bifurcation behaviors \red{(see bifurcating chemicals in Fig.~1 (5))}. These three arguments are generalized into multiple buffering structures, as stated in the introductory part of this paper. 
}

 \vspace{-0.3cm}
\begin{figure}[!htb]
\includegraphics[clip,width=3.8cm,trim=2 2 2 2]{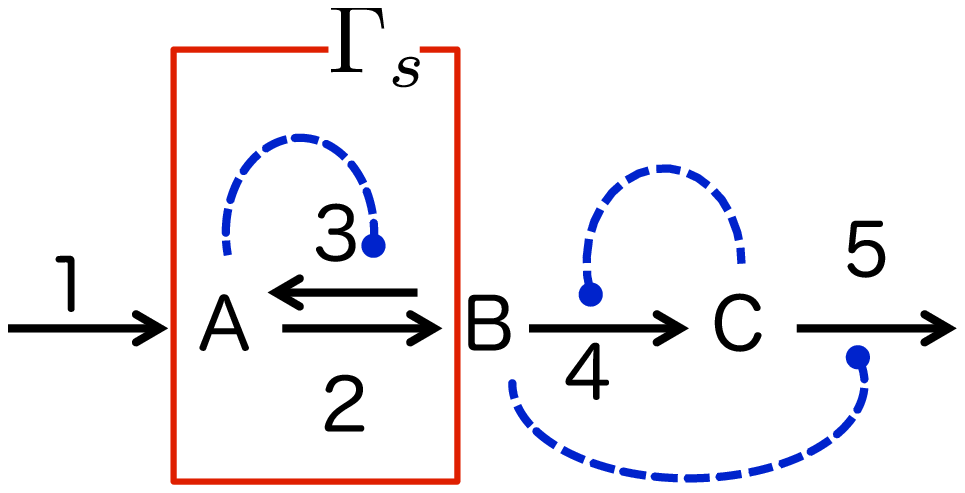}
\caption{ System consisting of five reactions, $\overset{k_1}{\rightarrow }{\rm A}, $ $\ {\rm A} \underset{k_2}{\overset{k_3}{\leftrightharpoons}} {\rm B}, \ {\rm B} \overset{k_4}{\rightarrow}{\rm C}, \ {\rm C} \overset{k_5}{\rightarrow}$. 
The three blue dashed arcs indicate {the regulations explained in the text}.
The red box indicates a buffering structure {$\Gamma_s$}. 
  }
\label{fig:sn_net}
\end{figure}

\section{ Hypothetical network}We demonstrate the structural bifurcation analysis in the system shown in Fig.~\ref{fig:sn_net}. 
The stoichiometry matrix $\nu$ and the kernel vectors are 
\bal
\nu &= \left(
\begin{array}{ccccc}
 1 & -1 & 1 & 0 & 0 \\
 0 & 1 & -1 & -1 & 0 \\
 0 & 0 & 0 & 1 & -1 \\
\end{array}
\right),\nonumber \\
{\mbf c}_1&= (0,1,1,0,0)^T,{\mbf c}_2 = (1,1,0,1,1)^T.
\eal

Every reaction rate depends on the substrate concentration. {Reaction rates} 
 $r_3,r_5$, and $r_4$ {are also}  regulated by A, B, and C, respectively. 
Then, ${\Gamma_s}=(\{ {\rm A}\}, \{ 2,3\})$ is a buffering structure since $\chi (\Gamma_s)=1-2+1=0$.

By permutating the row index  as $\{2,3,1,4,5 \}$ and the column index  as $\{\text{A},{\mbf c}_1,\text{ B},\text{ C}, {\mbf c}_2\}$, the matrix ${ \bf A}$ and the determinant are given by 
\bal
{ \bf A}=&
\left( 
\begin{array}{c|c}
{ \bf A}_{\Gamma_s} &{\bf  *}  \\ \hline
{\bf 0} & { \bf A}_{\bar{\Gamma}_s} 
\end{array}
\right)
= \left( 
\begin{array}{ccccc}
\cline{1-2}
\multicolumn{1}{|c}{ r_{2,\text{A}}} & \multicolumn{1}{c|}{ 1 }& 0 & 0 & 1 \\
\multicolumn{1}{|c}{ r_{3,\text{A}}}& \multicolumn{1}{c|}{1} & r_{3,\text{B}} & 0 & 0 \\
 \cline{1-5}
 0 & 0 & \multicolumn{1}{|c}{ 0} & 0 &  \multicolumn{1}{c|}{1} \\
 0 & 0 &\multicolumn{1}{|c}{  r_{4,\text{B}}} & r_{4,\text{C}} & \multicolumn{1}{c|}{ 1} \\
 0 & 0 &\multicolumn{1}{|c}{  r_{5,\text{B}} }& r_{5,\text{C}} &  \multicolumn{1}{c|}{1 }\\
  \cline{3-5}
\end{array}
\right),\nonumber \\
\detA &= \underset{\detA_{\Gamma_s}}{\underbrace{\left(r_{2,\text{A}}-r_{3,\text{A}}\right) }}\underset{\detA_{\bar{\Gamma}_s}}{\underbrace{\left(r_{5,\text{B}} r_{4,\text{C}}-r_{4,\text{B}} r_{5,\text{C}}\right)}}\label{detsn},
\eal
{where  nonzero  $\frac{\partial r_n}{\partial x_m}|_{{\mbf x}= {\mbf x^*}} $ is written by $r_{n,m}$, $m=$ A, B, C. } 
\corr{Due to the expression of $\det{\bf A}$, 
we conclude that the regulations corresponding to $r_{3,{\rm A} }$  and \red{$(r_{4,{\rm C}}, r_{5,{\rm B} })$}
are necessary for bifurcations associated with $\detA_\gs=0$ and $\detA_\gbs=0$, respectively. 
In this way, the possibility of bifurcation occurrences can be examined structurally  for each subnetwork. }

\begin{figure}[t]
\hspace{-0.2cm}
\includegraphics[clip,width=8.2cm,trim=1 2 0 0]{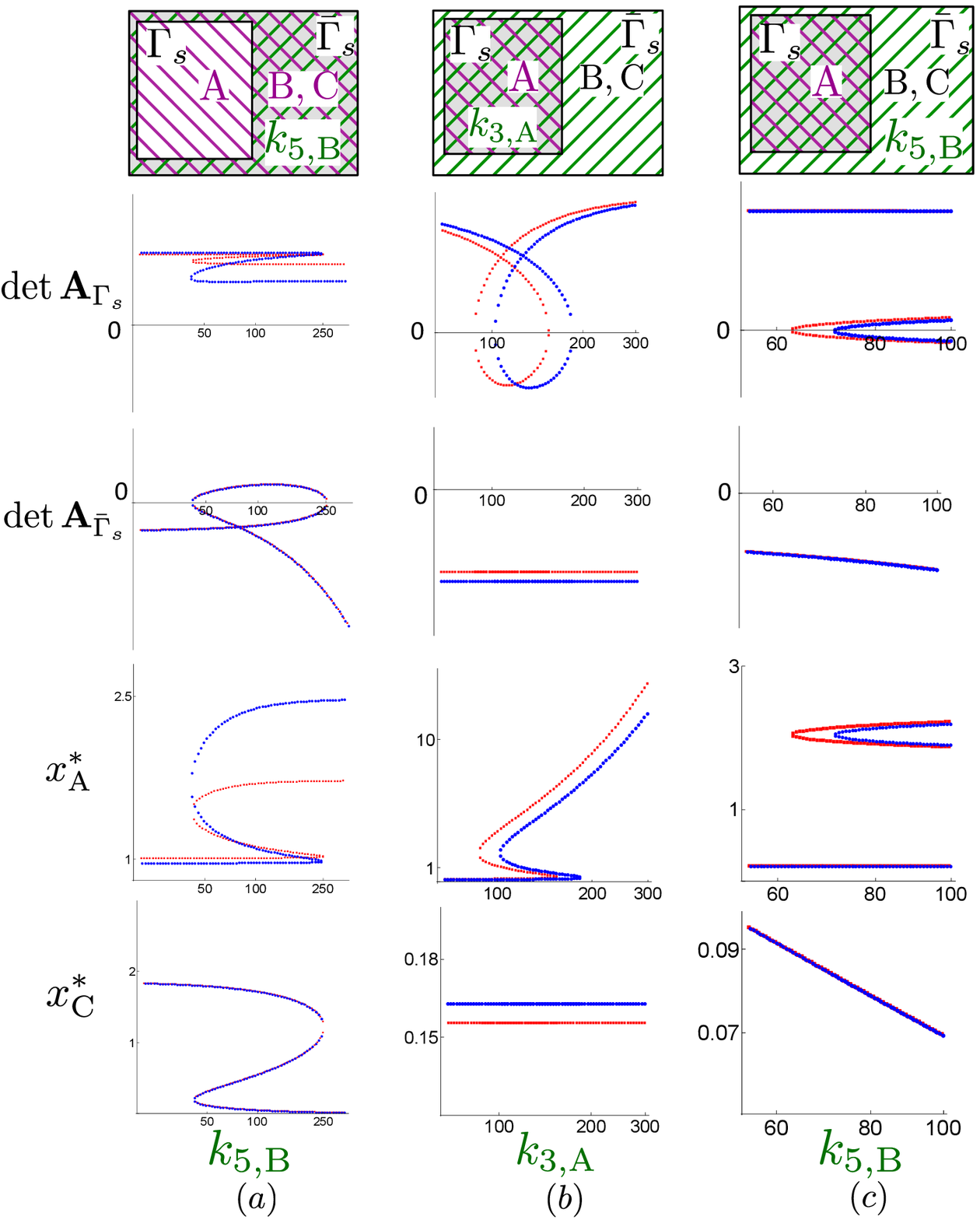}
\caption{{Steady-state values of $\det\, {\bf A}_{{\Gamma}_s}$ , $\det\, {\bf A}_{\bar{\Gamma}_s}$ , $x_A^*$, $x_C^*$ versus  $k_{5,{\rm B}}$ in (a), $k_{3,\rm A}$ in (b),  $k_{5,{\rm B}}$ in (c). }
\corr{Red and blue curves correspond to two choices of  parameters in $\gs$ in (a) and (c), two choices of parameters in $\gbs$  in (b).
In  the top panel of each case, the dark-colored region indicates the subnetwork whose determinant changes its sign  at the bifurcation point, and the  green- and purple-shaded regions indicate the inducing parameters and  bifurcating chemicals, respectively. 
See the SM for the specific parameter values used for the plots. }
} 
\label{fig:sn}
\end{figure}

For numerical demonstration, we assume the following kinetics:
\bal
{\mbf r} = &\biggl( k_1 , k_2 x_{\text{A}} ,k_3  x_{\text{B}}\left(1+ \frac{k_{3,\text{A}} x_{\text{A}}^2}{x_{\text{A}}^2+5}\right),\non \\
& k_4 x_{\text{B}} \left(1+ \frac{k_{4,\rm{C}} x_{\text{C}}^2}{x_{\text{C}}^2+5}\right) , k_5 x_{\text{C}}\left(1+ \frac{k_{5,\rm{B}} x_{\text{B}}^2}{x_{\text{B}}^2+5}\right)  \biggr).
\eal
\corr{The parameters are classified into $\{k_2, k_3, k_{3,A}\}\in \gs$, and  $\{k_1, k_4,k_5,k_{4,{\rm C}},k_{5,{\rm B}}\}\in \gbs$.}

\corr{First, we consider a bifurcation  associated with $\detA_{\bar\Gamma_s}$. The inducing parameters for the subnetwork $\gbs$  are the  parameters in ${\bar\Gamma}_s$ (the green-shaded region in the top panel of Fig.~3a). 
{As  seen from the plots of $\detA_{\Gamma_s}$ and $\detA_{\bar\Gamma_s}$ in Fig.~3a, the parameter  $k_{5,{\rm B}}$ in $\gbs$ actually induces  sign changes of $\detA_{\bar\Gamma_s}$ but not those of $\detA_{\Gamma_s}$ (two saddle-node bifurcations)}. The bifurcating chemicals for $\gbs$ are all chemicals $\{\text{A, B, C} \}$ (see the purple-shaded region of the top panel). Fig.~3a actually shows that both \red{chemicals} A and C exhibit steady-state bifurcations. 
On the other hand, varying  parameters in ${\Gamma_s}$ changes only concentrations in $\Gamma_s$, due to the law of localization. 
This is illustrated by  the blue and red curves in Fig.~3a.  
We   see that only the concentration of chemical A changes as  parameters in ${\Gamma_s}$ are varied, while the critical value of the bifurcation parameter  $k_{5,B}$ is independent \red{of parameters in} ${\Gamma_s}$.}

\corr{
Next, we consider the other bifurcation, which is associated with $\detA_{\Gamma_s}$.
The inducing parameters for the subnetwork $\gs$ consist of all  parameters in  $\gs$ and $\gbs$ (see the green-shaded region in the top \red{panels of Fig.~3b)}. 
The plots \red{ of $\detA_{\Gamma_s}$ in Fig.~3b show that} the  parameter $k_{3,{\rm A}}\in \Gamma_s$ indeed induces bifurcations associated with $\detA_\gs$. 
The bifurcating chemicals for $\gs$ are  the chemicals in $\gs$, i.e. $\{\text{A}\} $ (the purple-shaded region in the top panel). This can be confirmed from  \red{the plots for $x^*_{\rm A}, x^*_{\rm C}$} in Fig.~3b, where  only the steady-state of chemical A $\in \gs$ bifurcates at the bifurcation point, while chemical C $\in \gbs$ remains constant   as $k_{3,{\rm A}}\in \gs$ is varied, \green{due to} the law of localization. 
The law of localization also implies that 
varying \red{parameters  in ${\bar\Gamma}_s$ can}
 change concentrations of all chemicals. This is illustrated by the two curves in Fig.~3b{, where two different values of parameters in ${\gbs}$ result in  different concentration values for both of chemicals A and C.}
\red{In addition,} varying parameters in  $\gbs$ also influences the critical value of the bifurcation parameter $k_{3,\rm A}$. 
}

\corr{
There is another choice of bifurcation parameter  for the bifurcation associated with $\detA_{\Gamma_s}$,  as the inducing parameters for $\gs$ are not only parameters in $\gs$ but also those in $\gbs$  (see the green-shaded region in the top panel of Fig.~3c). 
 The parameter $k_{5,\rm B}\in \gbs$, which was  chosen as a bifurcation parameter in Fig.~3a, also induces the bifurcation associated with $\detA_{\Gamma_s}$ (see  the plots for $\detA_\gs$ in Fig.~3c).
This implies that the same parameter may induce bifurcations \red{for} different sets of chemicals depending on {which  factor 
of $\detA_{\Gamma_s}$ and $\detA_{\gbs}$ changes its sign}  at the critical points.
\textcolor{black}{As in the case of Fig.~3b, the bifurcating chemicals are in  $\Gamma_s$} (see the purple-shaded region \green{in}  the top panel of  Fig.~3c).
\textcolor{black}{Thus}  we  see that only chemical ${\rm A} \in \Gamma_s$ bifurcates at the critical point in Fig.~3c. \textcolor{black}{In particular, 
$k_{5,\rm B}$ does not induce the bifurcations of chemicals in $\gbs$, unlike the case of Fig.~3a.} On the other hand, because of  the law of localization, 
  parameters in  ${\Gamma_s}$
influence only  chemical ${\rm A} \in \gs$, as illustrated by the two curves in  Fig.~3c,  corresponding to  two different values of parameters in ${\gs}$. 
By contrast to Fig.~3a,
the critical value of the bifurcation parameter  $k_{5,B} \in {{\bar\Gamma}_s}$ is influenced by parameters in $\gs$.}

\begin{figure}[t]
\includegraphics[clip,width=7.5cm]{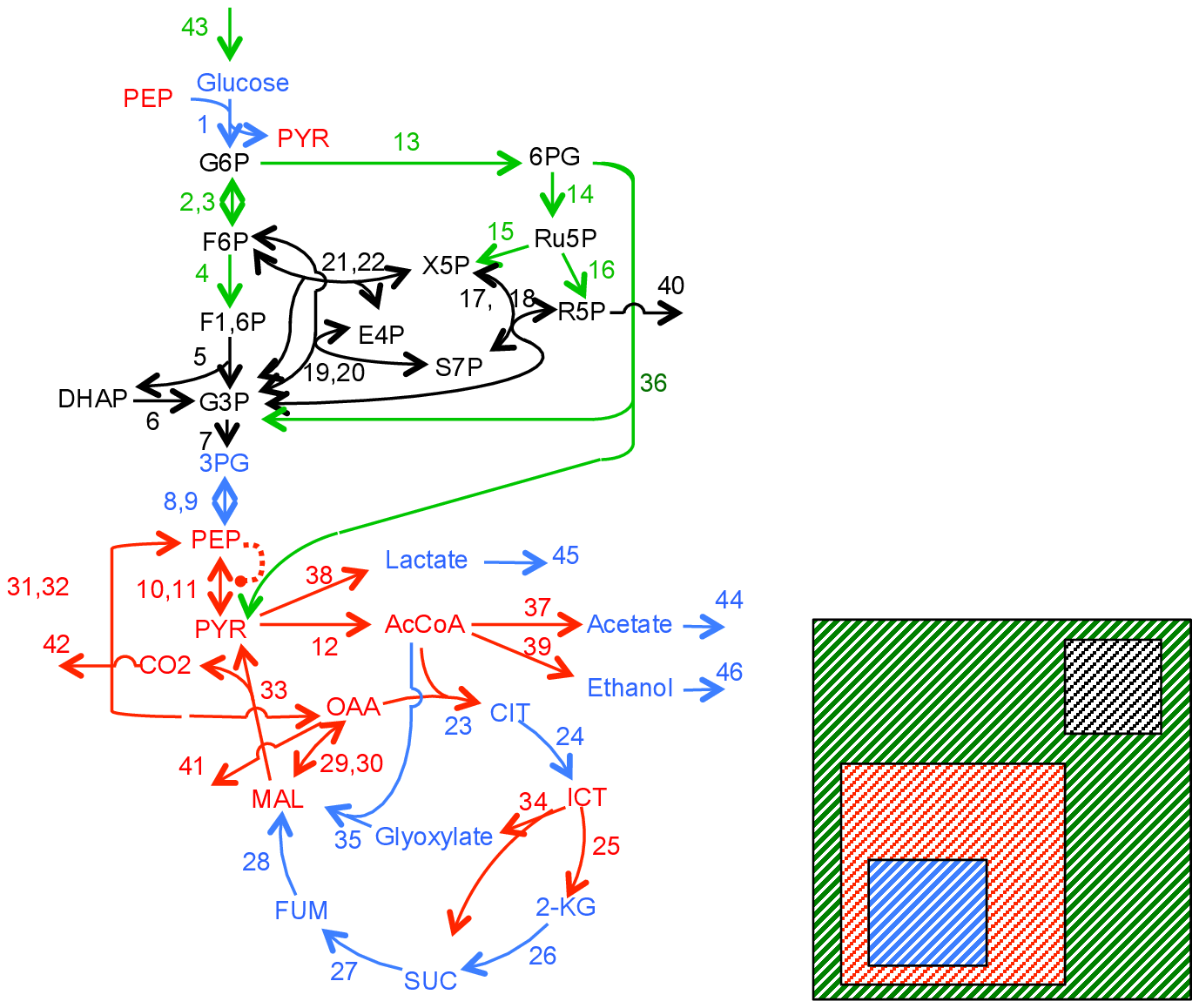}
\caption{{Central carbon metabolic network of E. coli. Dashed curve indicates a regulation from PEP to reaction 11 ($r_{11,{\rm PEP}}>0$). }　The right panel shows simplified inclusion relation of buffering structures. }
\label{fig:ec}
\end{figure}

{
\section{ E. coli network}\textcolor{black}{Finally, as a real example, we study the network of the
central carbon metabolism of E. coli \cite{OM} shown in Fig.~\ref{fig:ec}, consisting of 28 chemicals and 46 reactions. The network possesses 17 buffering structures in it.} 
 As shown in the SM, there is only \textcolor{black}{one} single subnetwork  (colored in red in Fig.~\ref{fig:ec}) that can satisfy the bifurcation condition. Reactions associated with the inducing parameters  \TO{for the red subnetwork} are colored in red and green. The bifurcating chemicals (and fluxes) are colored in red and blue. We confirmed this behavior numerically (see Fig.~S4 in the SM). 
}

\corr{
 Our method is applicable to any steady-state bifurcations of reaction systems even if systems have cokernel vectors. In the SM, we \red{ apply our theory} to a phosphorylation system \cite{CC} with cokernel vectors (i.e. conserved concentrations), and the First Schl\"ogl Model \cite{schlogl}, exhibiting transcritical bifurcations.
}

\corr{
\section{ Conclusion and perspectives}In this paper we propose a mathematical method to study bifurcation behaviors of  reaction systems from network structures alone.
In our method, bifurcations of \red{the} whole complex system are studied by factorizing it  into smaller substructures defined from buffering structures.
For each substructure, the bifurcation condition is studied, and a set of parameters possibly inducing onset of bifurcation is determined.
}

\corr{
Biological functions and their regulations are considered to arise from dynamics based on reaction networks and modulation of the enzymes.
On the other hand, the complexity of networks has been a large obstacle to understand such systems.
Our theory should be strongly effective to study dynamical behaviors of complex networks and promote systematic understandings of biological systems.
}

This work was supported partly by the CREST program (Grant No. JPMJCR13W6) of the Japan Science and Technology Agency (JST),  by iTHES Project/iTHEMS Program RIKEN, by Grant-in-Aid for Scientific Research on Innovative Area, Logics of Plant Development (Grant No. 25113001),  by NCTS and MOST of Taiwan. We express our sincere thanks to Gen Kurosawa, and Masashi Tachikawa for their helpful discussions and comments.

\appendix

\setcounter{equation}{0}
\setcounter{figure}{0}
\setcounter{table}{0}
\setcounter{page}{1}
\makeatletter
\renewcommand{\thefigure}{S\arabic{figure}}

\section*{\large \textbf{Supplemental Material}\ \\
Structural Bifurcation Analysis in Chemical Reaction Networks
}

\begin{center}
  Takashi Okada, Je-Chiang Tsai, and Atsushi Mochizuki\\
\end{center}

\section{Notation}
\begin{itemize}

\item
        $\mathbb X = \{X_1,\ldots, X_M\}$: the set of chemicals in the network.
        To ease the discussion, we also use the number index $m$ to identify the  $m$-th chemical $X_m$.
        The reader can distinguish this difference from the context.
\item
       $x_m$: the concentration of the $m$-th chemical, $m=1, \ldots, M$.
\item
         $\mathbb E = \{E_1,\ldots,E_N\}$: the set of  reactions in the network.
         Again, in order to ease the notations,
         we also use the number index to identify the reaction.
\item
       $r_n$: the reaction rate function associated with the $n$-th reaction, $n=1, \ldots, N$.

\item
       $k_n$: the  rate parameter associated with the reaction rate function  $r_n$, $n=1, \ldots, N$.
       Note that the reaction rate function $r_n$ is a function of $x_m$, $m=1,\ldots,M$.
       Here we assume that the $n$-th reaction rate function $r_n$ has only one rate parameter $k_n$ involved.
       But the analysis performed here indicates that we do not need this assumption.
       We do this just for the ease of presentation.
       Thus $r_n= r_n(\bx;k_n)$ where $\bx = (x_1,\ldots,x_M)^T$ and  $T$  denotes the transpose operator of a matrix.

\item ODE
\begin{align} \label{e:ode_app}
\frac{d x_m}{dt}= f_m(x_1,\ldots,x_M;k_1,\ldots,k_n) := \sum_{n=1}^N \nu_{mn} r_n(\bx;k_n),   \; m=1,\ldots,M.
\end{align}

\item  $\{{\mbf c}^\alpha\}_{\alpha=1}^K$ is the basis of the kernel space of the matrix $\nu = (\nu_{mn})_{M \times N}$.
$K:= {\rm dim \, ker}\, \nu$.
The steady state value $\br^*$ of the reaction   is given by a linear combination of them. We write
the coefficient of ${\mbf c}^\alpha$ by $\mu_\alpha $, i.e. $\br^* =\sum_{\alpha=1}^K  \mu_\alpha {\mbf c}^\alpha$.

\item $\{{\mbf d}_\beta\}_{\beta=1}^{K_c}$ is the basis of  the cokernel space of the matrix $\nu$
(i.e., ${\mbf d}_\beta^T\nu = 0$ for each $\beta = 1, \ldots,K_c$). $K_c:= {\rm dim \, coker}\, \nu$.
For each $ {\mbf d}_\beta$, using (\ref{e:ode_app})
the quantity $l_\beta \equiv{\mbf d}_\beta  \cdot  {\bx}$ is conserved with respect to time $t$.
\end{itemize}

Set
$$
    \bfe=(f_1,\ldots,f_M)^T,  \;  \br = (r_1,\ldots,r_N)^T,    \mbox{ and } \;
       {\mbf k} = (k_1,\ldots,k_N)^T.
$$
Then we can rewrite (\ref{e:ode_app}) as follows:
\be  \label{e:ode_short}
         \frac{d \bx}{dt} = \bfe(\bx;\mbf k) := \nu \br.
\ee

%

\section{The equivalence between the Jacobian matrix and the matrix {\bf A }}
In this section, we prove the equivalence between the Jacobian and the matrix {\bf A}. Systems with $K_c:= {\rm dim \, coker}\, \nu=0$ are discussed in section \ref{sec:eq0} (see \eqref{e:e:main_20_temp10_coker0}) and those with $K_c >0$ in section \ref{sec:eq>0} (see \eqref{e:e:main_20_temp10}).

\subsection{The proof of the equivalence when $K_c=0$}\label{sec:eq0}
To establish (\ref{e:e:main_20_temp10_coker0}),
we first note that the assumption $K_c =0$ implies that $N \ge M$ and $K=N-M$, where $K:= {\rm dim \ ker}\  \nu$. \footnote{In general, for any $M\times N$ matrix $\nu$, the identity $M+ {\rm dim \ ker} \ \nu = N + {\rm dim \ coker } \ \nu $ holds. }
Then employing the QR decomposition  to the matrix $\nu^T$,
we can factor the $\nu^T$
as the product of an $N \times N$ unitary matrix $\bQ$
and an $N \times M$ upper triangular matrix $\tilde{\bS}^T$
where $\tilde{\bS}^T$ is the transpose of the matrix $\tilde{\bS}$.
As the bottom $K$ rows of an $N \times M$ upper triangular matrix consist entirely of zeroes,
we thus have
\begin{align}  \label{e:nu_temp30}
   \nu^T  = \bQ \tilde{\bS}^T = [\bQ_1  \;  \bQ_2]
                  \left[\begin{array}{c}
                      \bS^T \\
                      {\bf 0}_{K \times M}\\
                    \end{array}\right]
                = \bQ_1  \bS^T,
\end{align}
where $\bS^T$ is an $M\times M$ upper triangular matrix, ${\bf 0}_{K \times M}$ is the $K \times M$ zero matrix,
$\bQ_1$ is a $N \times M$ matrix, $\bQ_2$ is a $N \times K$ matrix, and both $\bQ_1$ and $\bQ_2$ have orthogonal columns.
Further, the column space of $\bQ_1$ is equal to the row space of $\nu$,
whereas  the column space of $\bQ_2$ is equal to the null space of $\nu$,
which   follows from the fact
that the row space of any nontrivial matrix is the orthogonal complement to its null space.
To summarize, we have
$
   \nu =  \bS \bQ_1^T.
$
For simplicity, we let the basis $\{-\bc_\alpha\}_{\alpha=1}^K$ consist of the columns of $\bQ_2$.

Set
$
 \bw = [\bw_1 \;\;  \bw_2]^T
                    :=  \bQ^T \frac{\p \br}{\p  \bx},
$
where
$\bw_1$ is a $M \times M$ matrix, and $\bw_2$ is a $K \times M$ matrix.
Recall that $\bQ \bQ^T =I_N$.
Then we can rewrite the matrix $\frac{\p \br}{\p \bx}$ of $\br$ at the steady state $\bx^*$ as follows:
\be
 \frac{\p \br}{\p \bx} = \bQ (\bQ^T \frac{\p \br}{\p  \bx})  \\
     =  \big[\bQ_1  \;  \bQ_2\big]
      \left[\begin{array}{c}
                      \bw_1 \\
                      \bw_2\\
                    \end{array}\right] = \bQ_1\bw_1 + \bQ_2\bw_2.
\ee
Together with (\ref{e:nu_temp30}) and the orthogonal property of $\bQ$,
we can deduce that
\be
\begin{array}{ll}
 J =   \nu \dfrac{\p \br}{\p \bx}
         = (\bS \bQ_1^T) (\bQ_1\bw_1 + \bQ_2\bw_2)
         = \bS \bw_1.
\end{array}
\ee
This in turn implies
$
    \det(J) =  \det(\bS) \det(\bw_1).
$
Recall that the rank (rank($\nu$)) of the matrix $\nu$ is  $M$.
Then we have ${\rm rank}(\bS) = {\rm rank}(\bS \bQ_1^T) = {\rm rank}(\nu) = M$.
Hence $\bS$ is a square matrix of full rank,
and so we can conclude that
\be \label{c:df_coker0}
           \det(\bw_1) =  \big(\det(\bS)\big)^{-1} \det(J).
\ee

On the other hand, in view of the orthogonal properties of $\bQ, \bQ_1$, and $\bQ_2$,
the matrix {\bf A} (defined in the main text) can be written as
$$
   {\bf A} = [\frac{\p \br}{\p \bx} \; \bQ_2] = [\bQ\bw  \;\;   \bQ_2] = \bQ[\bw \;\; \bQ^T \bQ_2]
       =  \bQ \Big[ \begin{array}{l}
                       \bw_1 \\
                        \bw_2\\
                    \end{array}  \begin{array}{l}
                       {\bf 0}_{M \times K} \\
                        I_{K}\\
                    \end{array}\Big].
$$
Since  $\bQ$ is a square orthogonal matrix with $\det(\bQ) = \pm 1$,
we can thus deduce
\be  \label{c:A_coker0}
             \det({\bf w}_1)  = \pm \det({\bf A}).
\ee
%
Then from (\ref{c:df_coker0}) and (\ref{c:A_coker0}),
we obtain the following result.
\begin{theorem}
\be \label{e:e:main_20_temp10_coker0}
    \det({\bf A}) = \pm \frac{\det( J)}{\det(\bS)}.
\ee
\end{theorem}

\subsection{The proof of the equivalence when $K_c >0$}\label{sec:eq>0}
\subsubsection{The Jacobian matrix $\tJ$ for $K_c >0$} \label{sec:jac}

To study the stability of (\ref{e:ode_short}) around a steady state,
we need to compute the associated Jacobian matrix $J = D_\bx \bfe = \nu \frac{\p \br}{\p \bx}$.
On the other hand,
when  \red{$K_c >0$},
there are only $\tM \equiv M- K_c$ independent variables in the set $\{x_m\}_{m=1}^M$.
Thus  we need to eliminate $K_c$ redundant variables from the set $\{x_m\}_{m=1}^M$
before computing the Jacobian matrix $J$.

Indeed, when \red{$K_c >0$},
without loss of generality,
we may assume that the first $K_c$ rows of $\nu$ are linearly dependent.
Then we can rewrite $\nu$ in the following form \cite{Klipp},
\bal
\nu = \left(
\begin{array}{c}
\tnu\\
 L \tnu
\end{array}\right) =\left(
\begin{array}{c}
I_{\tM} \\
 L
\end{array}\right) \tnu, \label{nudecompose}
\eal
where
$\tnu$ is  a $\tM \times N$ matrix, and  $L$ (called a {\it link matrix}) is a $K_c \times \tM$ matrix.
Here the matrix $I_{\tM} $ is the $\tM \times \tM$ identity matrix.
In the remaining of this appendix,
the terminology for the $n \times n$ identity matrix $I_n$  with $n\in\bN$ is employed.

Decompose $\bx \in \bR^M$ as follows:
$$
    \bx =
\left(
\begin{array}{c}
{\mbf x}_{ind}\\
{\mbf x}_{dep}
\end{array}
\right),
$$
where
${\mbf x}_{ind}\in {\mathbb R}^{\tM}$ and ${\mbf x}_{dep}\in {\mathbb R}^{K_c}$.
With these notations defined as above,
we can rewrite the ODE system~(\ref{e:ode_app}) (or (\ref{e:ode_short}))  in the following form,
\bal
 \frac{d \bx}{dt} =
\frac{d}{dt}\left(
\begin{array}{c}
{\mbf x}_{ind}\\
{\mbf x}_{dep}
\end{array}
\right)= \left(
\begin{array}{c}
I_\tM\\
L
\end{array}
\right)\tnu \red{\br}. \label{ode_sep}
\eal
Note $\frac{d}{dt}{\mbf x}_{dep} =  L \tnu  {\mbf r}  = L \frac{d}{dt}{\mbf x}_{ind}$.
Then an integration of this identity indicates that
the ODE system (\ref{ode_sep}) is equivalent to the following system:
\bal
\frac{d}{dt}{\mbf x}_{ind} &= \tnu \red{\br}, \label{indep}\\
-  L {\mbf x}_{ind} + {\mbf x}_{dep}  &= constant. \label{const}
\eal

Now we compute the Jacobian matrix $\tJ$ associated with the ODE system (\ref{indep}).
To proceed, let $\bx^* = (\bx^*_{ind}, \bx^*_{dep})$ be a steady state of (\ref{ode_sep}).
Note that the reaction function $\br$ depends not only on ${\mbf x}_{ind}$, but also on ${\mbf x}_{dep}$.
Then the Jacobian matrix $\tJ$ associated with the ODE system (\ref{indep}) around $\bx^*_{ind}$ is given by
\be \label{J'0}
\begin{array}{ll}
\tJ    = \tnu  \dfrac{d \br  }{d {\mbf x_{ind}}}
        & := \tnu  \Big(\dfrac{\p \br }{\p {\mbf x}_{ind}}
                        +\dfrac{\p \br }{\p {\mbf x}_{dep}} \cdot \dfrac{\p \bx_{dep} }{\p {\mbf x}_{ind}} \Big)\biggr|_{\bx = \bx^*}   \\[2.5ex]
     & = \tnu  \Big(\dfrac{\p \br }{\p {\mbf x}_{ind}}+\dfrac{\p \br }{\p {\mbf x}_{dep}}L \Big)\biggr|_{\bx = \bx^*}  \\[2.5ex]
     & = \tnu  \Big(\dfrac{\p \br }{\p {\mbf x}_{ind}}+\dfrac{\p \br }{\p {\mbf x}_{dep}} \Big)\biggr|_{\bx = \bx^*}
            \cdot \left(
            \begin{array}{c}
              I_{\tM}\\
               L
             \end{array}
             \right)
             \end{array}
             \ee
Note that the size of $\tJ$ is $\tM \times \tM$, that of $\frac{\p \br }{\p {\mbf x}_{ind}}$ is $N\times \tM$,
and that of $\frac{\p \br }{\p {\mbf x}_{dep}}$ is $N \times K_c$.

\subsubsection{QR decomposition of the stoichiometric matrix $\nu$ for $K_c >0$}

Consider the QR decomposition of the $\tM \times N$ matrix $\tnu$.
First, since  $\tnu$ is of full rank, we have $N \ge \tM$.
Then employing the QR decomposition  to the matrix $\tnu^T$,
we can factor the $\tnu^T$
as the product of an $N \times N$ unitary matrix $\bQ$
and an $N \times \tM$ upper triangular matrix $\tilde{\bS}^T$
where $\tilde{\bS}^T$ is the transpose of the matrix $\bS$.
As the bottom $(N - \tM)$ rows of an $N \times \tM$ upper triangular matrix consist entirely of zeroes,
we thus have
$$
   \tnu^T  = \bQ \tilde{\bS}^T = [\bQ_1  \;  \bQ_2]
                  \left[\begin{array}{c}
                      \bS^T \\
                      {\bf 0}_{(N - \tM) \times \tM}\\
                    \end{array}\right]
                = \bQ_1  \bS^T,
$$
where $\bS^T$ is an $\tM\times \tM$ upper triangular matrix, ${\bf 0}_{(N - \tM) \times \tM}$ is the $(N - \tM) \times \tM$ zero matrix,
$\bQ_1$ is a $N \times \tM$ matrix, $\bQ_2$ is a $N \times (N - \tM)$ matrix, and both $\bQ_1$ and $\bQ_2$ have orthogonal columns.
Further, the column space of $\bQ_1$ is equal to the row space of $\tnu$,
whereas  the column space of $\bQ_2$ is equal to the null space of $\tnu$.
The second statement  follows from the fact
that the row space of any nontrivial matrix is the orthogonal complement to its null space.
To summarize, we have
\be
   \tnu =  \bS \bQ_1^T.
\ee

\subsubsection{The  matrix {\bf A} for $K_c>0$ (review of \cite{OMpre})} \label{sec:app_1.3}
We review the definition of the matrix {\bf A}  for $K_c>0$ \cite{OMpre} and express it by using the bases of the kernel and cokernel spaces of $\nu$, introduced in the previous two sections.
For a chemical reaction system with $K_c  >0$,  the steady state depends continuously on the reaction rate $k_n (n=1,\ldots, N)$  and also on  the initial values  of conserved concentrations, $l_\beta (\beta=1,\ldots, K_c)$. Thus,  we consider $N + K_c$ perturbations, in total. The sensitivity to the $N + K_c$ perturbations can be determined at once by solving the following equation,
\begin{align}
{\bf A} \left(
\begin{array}{cccc}
\delta_1 {  \mbf x} &  \ldots  &\delta_{N + K_c} { \mbf  x}  \\\hline
\delta_{1} {  \mbf \mu} &  \ldots & \delta_{N + K_c} {  \mbf \mu}  \\
\end{array}
\right)
= -   \left(
\begin{array}{c| c}
{\bf E}_{N}&{\bf  0}\\\hline
{\bf 0}& {\bf E'}_{K_c}
\end{array}
\right),\label{Adx=-E}
\end{align}
where
${\bf E}_{N}$  and  ${\bf E'}_{K_c}$ are  $N \times N$ and $K_c \times K_c$ diagonal matrices, respectively,
whose components are proportional to the perturbations;
the $n$-th  component of ${\bf E}_{N}$ is given by $\frac{\partial  r_n}{\partial k_n} \delta k_n$,
and the $\beta$-th component of ${\bf E}_{K_c}$ is given by  $\delta l_\beta$.
The matrix ${\bf A}$ is given by
\begin{align}
{\bf A}  :=  \begin{array}{c}
\\
\ _N \Bigg{\updownarrow} \\
\\
\ _{K_c}\Bigg{\updownarrow}\\
\\
\end{array}
\underset{\underset{M}{\xlongleftrightarrow[]{\hspace{4em}}} \ \ \ \ \ \ \underset{K}{\xlongleftrightarrow[]{\hspace{5em}}}}{\left(
\begin{array}{cccc|c }
&&&&  \\
&\mbox{\smash{ $\dfrac{\p \br  }{\p {\mbf x}}$}}&&&{ - {\mbf c}^{\,1}\ {\ldots}\ - {\mbf c}^{\,K}}\\
&&&&\\\hline
&{-({\mbf d}_{\,1})^T}& & &\\
&\vdots&&&\mbox{\smash{\ ${\bf 0}_{K_c \times K}$}} \\
&{-({\mbf d}_{\,K_{c}})^T}& & &
\end{array}
\right)}, \label{Amat}
\end{align}
where ${\bf 0}_{K_c \times K}$ is a $K_c \times K$ zero matrix.
With a direct computation, one can see that
the row vectors of the following $K_c \times M$ matrix,
\bal
D := (-L \;\;  I_{K_c})
\eal
spans the cokernel space of $\nu$.
Recall that $\{{\mbf d}_\beta\}_{\beta=1}^{K_c}$ is the basis of  the cokernel space of the matrix $\nu$.
To facilitate the computation, we set
$$
      \left(
     \begin{array}{c}
       -({\mbf d}_{\,1})^T \\
       \vdots \\
       -({\mbf d}_{\,K_c})^T
      \end{array}
      \right) = D.
$$
Also recall that $\{{\mbf c}^\alpha\}_{\alpha=1}^{K}$ is the basis of  the kernel $ker(\nu)$ of the matrix $\nu$,
and that  the column space of $\bQ_2$ is  the kernel $ker(\tnu)$ of the matrix $\tnu$.
One can show that $ker(\nu) = ker(\tnu)$.
Motivated by this, we set
$$
      (-{\mbf c}^1,\ldots,-{\mbf c}^K) = \bQ_2.
$$
Taken together, the  matrix $\bf A$ takes the following form:
\begin{align}
{\bf A} = \begin{array}{c}
\\
\ _N\Bigg{\updownarrow} \\
\ _{K_c}\bigg{\updownarrow}\\
\end{array}
\underset{\underset{M}{\xlongleftrightarrow[]{\hspace{8em}}} \ \ \ \ \ \ \underset{K}{\xlongleftrightarrow[]{\hspace{3em}}}}{\left(
\begin{array}{cccc|c }
&&&&  \\
&\mbox{\smash{ $ \dfrac{\p \br  }{\p {\mbf x_{ind}}}$ \  $\dfrac{\p \br }{\p {\mbf x_{dep}}}$}}&&& { \bQ_2} \\
&&&&\\\hline
& -L   \hspace{1.2 cm}  I_{K_c}    &&&\mbox{\smash{\ ${\bf 0}_{K_c \times K}$}} \\
\end{array}
\right)}. \label{Amat2}
\end{align}

\subsubsection{The relation between the Jacobian matrix $\tJ$ and the matrix {\bf A}  for $K_c >0$}  \label{sec:app_1.4}

Set
\be \label{e:bw}
    \bw = \left[\begin{array}{c}
                      \bw_1 \\
                      \bw_2\\
                    \end{array}\right]
                    :=  \bQ^T \frac{d \br}{d \bx_{ind}} \\
                    = \bQ^T  (\frac{\p \br }{\p {\mbf x}_{ind}}+\frac{\p \br }{\p {\mbf x}_{dep}}L),
\ee
where
$\bw_1$ is a $\tM \times \tM$ matrix, and $\bw_2$ is a $ (N - \tM) \times \tM$ matrix.
Recall that $\bQ \bQ^T =I_N$.
Then we can rewrite the  matrix $ \frac{d \br}{d \bx_{ind}} $ as follows:
\be
\begin{array}{ll}
   \dfrac{d \br}{d \bx_{ind}}  &= \bQ (\bQ^T \dfrac{d \br}{d \bx_{ind}}),  \\ [2ex]
 &
     =  \big[\bQ_1  \;  \bQ_2\big]
      \left[\begin{array}{c}
                      \bw_1 \\
                      \bw_2\\
                    \end{array}\right] = \bQ_1\bw_1 + \bQ_2\bw_2.
\end{array}
\ee
Recall that $\tnu = \bS \bQ_1^T$.
Together with the fact that $\bQ_1^T \bQ_1 =I_\tM$
and $\bQ_1^T \bQ_2  = {\bf 0}_{\tM \times (N - \tM)}$
where ${\bf 0}_{\tM \times (N - \tM)}$ is the $\tM \times (N - \tM)$ zero matrix,
we can deduce that
\be
\begin{array}{ll}
 \tJ =   \tnu \dfrac{d \br}{d \bx_{ind}}  &= (\bS \bQ_1^T) (\bQ\bw)  \\
         &= (\bS \bQ_1^T) (\bQ_1\bw_1 + \bQ_2\bw_2). \\
         &= \bS \bw_1.
\end{array}
\ee
This in turn implies
$$
    \det(\tJ) =  \det(\bS) \det(\bw_1).
$$
Recall that the rank of the matrix $\tnu$ is $rank(\tnu) = \tM$.
Then we have $rank(\bS) = rank(\bS \bQ_1^T) = rank(\tnu) = \tM$.
Hence $\bS$ is a square matrix of full rank.
Thus we can conclude that
\be \label{c:df}
           \det(\bw_1) = \frac{\det(\tJ)}{\det(\bS)}.
\ee
%

On the other hand, using (\ref{e:bw}) we have
\begin{align*}
{\bf A} =
\left(
\begin{array}{cccc|c }
&&&&  \\
&\mbox{\smash{ $ \dfrac{\p \br  }{\p {\mbf x_{ind}}}$   \quad  $\dfrac{\p \br }{\p {\mbf x_{dep}}}$}}&&&{ \bQ_2  }\\
&&&& \\\hline
&-L   \hspace{1.2 cm}  I_{K_c} &&&\mbox{\smash{\ ${\bf 0}_{K_c \times K}$}} \\
\end{array}
\right)
=
\left(
\begin{array}{cccc|c }
&&&&  \\
&\mbox{\smash{ $\bQ\bw -  \dfrac{\p \br  }{\p {\mbf x_{dep}}}L$   \quad  $\dfrac{\p \br }{\p {\mbf x_{dep}}}$}}&&&{ \bQ_2  }\\
&&&& \\\hline
&-L   \hspace{2 cm}  I_{K_c} &&&\mbox{\smash{\ ${\bf 0}_{K_c \times K}$}} \\
\end{array}
\right). 
\end{align*}
To compute $\det({\bf A})$,
consider the matrix
\begin{align*}
{\bf B} =
\left(
\begin{array}{cccc|c }
&&&&  \\
&\mbox{\smash{ $\bQ\bw$   \quad  $\bQ_2$}}
            &&& \dfrac{\p \br }{\p {\mbf x_{ind}}}\\
&&&& \\\hline
& {\bf 0}_{K_c \times \tM}    \hspace{1 cm}  {\bf 0}_{K_c} &&&\mbox{\smash{\ $I_{K_c}$}} \\
\end{array}
\right). 
\end{align*}
It is verified that $\det({\bf A}) = (-1)^{K_c(K_c + 1)/2} \det({\bf B})$.
Thus it suffices to consider the matrix
$$
   {\bf C} := [\bQ\bw  \;\;   \bQ_2].
$$
In view of the fact that $\bQ^T \bQ = I_N$,
$\bQ_1^T \bQ_2  = {\bf 0}_{\tM \times (N-\tM)}$,
and $\bQ_2^T \bQ_2 =I_{N-\tM}$,
we have
$$
   \bQ^T {\bf C} = [\bQ^T \bQ \bw  \;\; \bQ^T \bQ_2]
                 = \Big[\bw  \begin{array}{l}
                       {\bf 0}_{M \times (N - \tM)} \\
                        I_{N - \tM}\\
                    \end{array}\Big]
                   =  \Big[ \begin{array}{l}
                       \bw_1 \\
                        \bw_2\\
                    \end{array}  \begin{array}{l}
                       {\bf 0}_{M \times (N - \tM)} \\
                        I_{N - \tM}\\
                    \end{array}\Big].
$$
This in turn implies
\be
    {\bf C} = \bQ \Big[ \begin{array}{l}
                       \bw_1 \\
                        \bw_2\\
                    \end{array}  \begin{array}{l}
                       {\bf 0}_{M \times (N - \tM)} \\
                        I_{N - \tM}\\
                    \end{array}\Big].
\ee
Since  $\bQ$ is a square orthogonal matrix with $\det(\bQ) = \pm 1$,
we can thus deduce
\be  \label{c:A}
             \det({\bf w}_1)  = \pm \det({\bf A}).
\ee
%

Then from (\ref{c:df}) and (\ref{c:A}),
we can conclude the following theorem.
\begin{theorem} \label{t:main_20}
The following equality holds.
\be \label{e:e:main_20_temp10}
    \det({\bf A}) = \pm \frac{\det(\tJ)}{\det(\bS)}.
\ee
\end{theorem}



\section{ Structural Factorization of det {\bf A} }
Here, we explain the factorization of {\bf A} in detail, after reviewing buffering structures and the law of localization.
Systems with  $K_c =0$ are discussed in section \ref{sec:fac0}, and those with  $K_c >0$ in
section \ref{sec:fac}.

\subsection{ Structural Factorization when $K_c =0$}\label{sec:fac0}

\subsubsection{Buffering structure and the law of localization when $K_c =0$ (review of \cite{OM})}

We construct a {\it subnetwork} $\Gamma_s = (\mathfrak m, \mathfrak n)$, a pair of chemicals and reactions, as follows:
\begin{enumerate}
\item Choose a subset $\mathfrak m \subseteq \mathbb X$ of chemicals.
\item Choose a subset $\mathfrak n \subseteq \mathbb E$ of reactions such that $\mathfrak n$  includes all reactions
whose reaction rate functions $r_n(\mbf{x})$ depend on at least one member in $\mathfrak m$. Namely, we can construct $\mathfrak n$ by collecting all reactions $n$ that are regulated by  $\mathfrak m$
\footnote{The phrase that the reaction $n$ is regulated by the chemical $m$ means that $\frac{\p r_n}{\p x_m} \not = 0$.}
 plus any other reactions.
\end{enumerate}
Below, we consider only subnetworks constructed in this way and  call them  {\em subnetworks}.
To proceed, we introduce the definition that a vector ${\bf v} \in\bR^{N}$ has support contained in $\mathfrak n$.
Indeed, for the reaction subset ${\mathfrak n}$,  we can associate the vector space $V({\mathfrak n})$:
$$
V({\mathfrak n})  :=   \, {\rm span}\, \bigl\{    {\bf v}   | \,  {\bf v} \in {\rm ker}\, \nu , P^{\mathfrak n}  {\bf v} =  {\bf v} \bigr\}.
$$
Here, $P^{\mathfrak n}$ is an $N \times N$ projection matrix onto the space associated with ${\mathfrak n}$ defined by
$$
P^{\mathfrak n}_{n,n'} = \delta_{n,n'}\   {\rm if } \  n,n' \in {\mathfrak n}.\  {\rm Otherwise } \ P^{\mathfrak n}_{n,n'} =0.
$$
Then we say that a vector ${\bf v} \in\bR^{N}$ has support contained in $\mathfrak n$
if ${\bf v} \in  V({\mathfrak n})$.

For a subnetwork $\Gamma_s = (\mathfrak m, \mathfrak n)$, we define the {\it index} $\chi(\Gamma_s)$ by the relation
\bal
\chi(\Gamma_s) = |\mathfrak m| -| \mathfrak n| + (\# cycle).
\eal
Here, $|\mathfrak m|$ and  $| \mathfrak n|$ are the number of elements in the sets $\mathfrak m$ and  $\mathfrak n$, respectively.
The $\# cycle$ is the number of independent  kernel vectors of the matrix $\nu$ whose supports are contained in $\mathfrak n$.
In general, $\chi(\Gamma_s)$ is non-positive (see \cite{OM}).

Then a {\it buffering structure} is defined as a subnetwork $\Gamma_s$ with $\chi(\Gamma_s) =0$.

%


It was proved in \cite{OM} that, for a buffering structure  $\Gamma_s = (\mathfrak m, \mathfrak n)$,
 the steady state values of chemical concentrations and reaction rates  outside $\Gamma_s$ are independent of the reaction rate parameters $k_n$
 of reactions in $\mathfrak n$.
 Specifically, let $\bx^*=(x_1^*,\ldots,x_M^*)$ be the steady state of (\ref{e:ode_app}).
 Note that $\bx^*$ depends on the parameter vector ${\mbf k}=(k_1,\ldots,k_N)^T$.
Set
$$
\br^*=(r_1^*,\ldots,r_N^*):=(r_1(\bx^*;k_1),\ldots,r_N(\bx^*;k_N)).
$$
Then, for any $n'  \in  \mathfrak n$, and any $m \in \mathfrak m^c$
and $n \in \mathfrak n^c$,
one has
\bal
\frac{\p x_m^*}{\p k_{n'}} = 0, \;\;   \frac{\p  r_{n}^*}{\p k_{n'}} = 0, \label{lol}
\eal
where $\mathfrak m^c = \mathbb X \setminus \mathfrak m$
and $\mathfrak n^c = \mathbb E \setminus \mathfrak n$
are the complementary set of $\mathfrak m$ and $\mathfrak n$, respectively.
For ease of notation, we set
$$
       \bar \Gamma_s :=  (\mathfrak m^c, \mathfrak n^c).
$$

We remark that a whole network $\Gamma$ always satisfies the condition of the law of localization because $\chi(\Gamma) =M -N + K = K_c =0$.


\subsubsection{Factorization of the matrix  {\bf A} when $K_c =0$}
Suppose that $\Gamma_s$ is a buffering structure.
Then, by permuting  the columns and rows of the matrix {\bf A},
the resulting matrix (still denoted by $\bf A$ for simplicity)  can be written in the following form \cite{OM}:
\bal
 {\bf A} =\begin{pmatrix}
 {\bf A}_{\Gamma_s} & {\bf A}_{\Gamma_s,\bar \Gamma_s} \\
 {\bf 0}_{ |\mathfrak n^c|\times |\mathfrak n| } & {\bf A}_{\bar \Gamma_s}
 \end{pmatrix}. \label{Amat_coker0}
\eal

The entries of $\green{{\bf A}_{\Gamma_s}}$ consist of components of  the constant vectors ${\bf v} \in V(\mathfrak n)$,
and the term $\frac{\p r_n}{\p x_m}\big|_{\bx=\bx^*}$ with $(m, n) \in \Gamma_s$ \footnote{For ease of presentation,  for a  subnetwork $\Gamma_s = (\mathfrak m, \mathfrak n)$, if no confusions can arise, we write $(m, n) \in \Gamma_s$  instead of ``$m \in \mathfrak m$ and $n \in \mathfrak n$''.}
which  corresponds to  self-regulations inside $\Gamma_s$
(regulations from chemicals in $\Gamma_s$ to reaction rate functions in $\Gamma_s$).
Similar characteristics for the entries of  ${\bf A}_{\bar \Gamma_s}$ can be observed,
which corresponds to self-regulations inside the complementary part $\bar \Gamma_s$.
The non-constant entries of the upper-right matrix ${\bf A}_{\Gamma_s,\bar \Gamma_s} $
\red{correspond} to regulations from  $\bar \Gamma_s$ to  $\Gamma_s$. \corr{Finally,  the lower-left block   in \eqref{Amat_coker0}  is the zero matrix because (i) the kernel vectors in $\gs$ do not have support on reactions in $\mathfrak n^c$ and (ii) $\partial r_n/\partial x_m =0$ for $n \in \mathfrak n^c$, $m\in \mathfrak m$ follows from the condition that $\gs$ is output-complete (see also \cite{OM})}. 

Therefore, we have the following results.
\begin{theorem}
The following hold for a buffering structure $\Gamma_s$ in a network $\Gamma$.
\begin{enumerate}
\item[\rm(i)]
The determinant of the  matrix $\bf A$ can be factorized as follows:
\bal
\det {\bf A} = \det {\bf A}_{\Gamma_s} \times \det {\bf A}_{\bar \Gamma_s}. \label{factorize}
\eal
Note that ${\bf A}_{\Gamma_s, \bar \Gamma_s} $ does not contribute to  $\det {\bf A}$.
\item[\rm(ii)]
\bal
\frac{\p {\bf A}_{\bar \Gamma_s}}{\p k_n}=0  \ \  {\rm for} \   n\in \mathfrak n. \label{detAind}
\eal
Thus the complementary part ${\bf A}_{\bar \Gamma_s}$ is independent of the rate parameters $k_n $
with $n\in \mathfrak n$.
\end{enumerate}
\end{theorem}

The assertion (ii) can be proved as follows:
The entries of ${\bf A}_{\bar \Gamma_s}$ consist of the term $\frac{\p  r_{n'}}{\p x_{m'}}\big|_{\bx=\bx^*}$ with $(m',n') \in (\mathfrak m^c,\mathfrak n^c)$ and the components of the kernel vectors of the matrix $\nu$,
which is obviously independent of $k_n$ for each $n=1,\ldots,N$. 
From the construction of the subnetwork $\Gamma_s$, $r_{n'}$ are functions of variables $x_{m'}$ with $m'\in \mathfrak m^c$ \footnote{If the function $r_{n'}$ depended also on $x_m$ with $m\in \mathfrak m$, such a reaction $n'$ should be included into $\Gamma_s$, by construction of the subnetwork $\Gamma_s$}, and so  the derivatives $\frac{\p r_{n'}}{\p x_{m'}}\big|_{\bx=\bx^*}$ can be written in terms of  $x_{m'}^*$ with $m'\in \mathfrak m^c$. Then, these derivatives $\frac{\p r_{n'}}{\p x_{m'}}\big|_{\bx=\bx^*}$ are independent of $k_n$ for $n \in \mathfrak n$ due to \eqref{lol}.

\corr{An important remark is that the above theorem can  be applied not only to a single buffering structure  but also for nested buffering structures. For example, a buffering structure  within another larger buffering structure is studied similarly}, by regarding $\Gamma$  as a larger buffering structure and $\Gamma_s$ as a smaller buffering structure in it.\footnote{Recall that a whole network $\Gamma$ is always a buffering structure.}

In summary, for  a buffering structure $\Gamma_s$ inside  a network $\Gamma$, we have proved
\bal
\det {\bf A}_\Gamma = \det {\bf A}_{\Gamma_s}  \times \underset{
\underset{ k_{\Gamma_{s}}}{{\rm independent \  of \  } }
}{\underbrace{\det {\bf A}_{\bar \Gamma_{s}}} },\label{factorize'}
\eal
where $k_{\Gamma_s}$ denotes the set of parameters of reactions inside $\Gamma_s$. 

%
%

\subsubsection{Factorization for multiple buffering structures when $K_c =0$}
We generalize the above factorization formula \eqref{factorize'}  into multiple buffering structures.

First, we consider a network  $\Gamma$ containing  $L$ non-intersecting buffering structures $\Gamma_1,\ldots, \Gamma_L$. We write  the complement of the buffering structures as $\bar \Gamma_{1,\ldots,L} :=\Gamma\backslash (\Gamma_1\cup \ldots \cup \Gamma_L) $;
\bal
\Gamma= \Gamma_1\cup \ldots \cup \Gamma_L \cup {\bar \Gamma}_{1,\ldots,L}.
\eal
In this case, the matrix {\bf A} can be written in the following form\footnote{\corr{For each  structure $\gs$, the columns associated with chemicals and cycles in $\gs$ have nonzero entries only for  reactions in $\gs$, \green{by} the same reason explained below  \eqref{Amat_coker0}. Thus, zero matrices appear in the non-diagonal blocks in \eqref{Awointer}. }};
\bal
{\bf A} =
\left(\begin{array}{ccccc}
\cellcolor[gray]{.8} {\bf A}_{\Gamma_1} & {\mbf 0} & \cdots &{\mbf 0}& {\bf A}_{\Gamma_1, \bar \Gamma_{1,\ldots,L}}  \\
  {\mbf 0} &\cellcolor[gray]{.8} {\bf A}_{\Gamma_2}  & \cdots &{\mbf 0} &  {\bf A}_{\Gamma_2, \bar \Gamma_{1,\ldots,L}}\\
  \vdots  & \vdots  & \ddots & \vdots &\vdots \\
{\mbf 0}& {\mbf 0} & \cdots &  \cellcolor[gray]{.8}{\bf A}_{\Gamma_L} & {\bf A}_{\Gamma_L, \bar \Gamma_{1,\ldots,L}}\\
{\mbf 0}& {\mbf 0}& \cdots &{\mbf 0}&  {\bf A}_{\bar \Gamma_{1,\ldots,L}}
 \end{array}\right),\label{Awointer}
\eal
where each shaded block ${\bf A}_{\Gamma_s}$ ($s=1,\ldots,L$) is an square matrix associated with $\gs$, defined as
\bal
{\bf A}_{\Gamma_s} =
\left(\begin{array}{c|ccc}
({\frac{ {\mbf \partial{\mbf  r}}}{{\mbf \partial}{\mbf x}}})_{\Gamma_s} &{\mbf c^s}_1 & \ldots & {\mbf c^s}_{K_s}
 \end{array}\right).
\eal
Here, the left block, $({\frac{ {\mbf \partial{\mbf  r}}}{{\mbf \partial}{\mbf x}}})_{\Gamma_s}$, consists of $\frac{\partial r_n}{\partial x_m}$ with $(m,n)\in \Gamma_s$, and the right block corresponds to the  independent kernel vectors of $\nu$ whose support are on reactions in $\Gamma_s$ ($K_s$ is the number of the  kernel vectors).

Then, we can prove that
\bal
\det {\bf A} = \biggl (\prod_{s=1}^L \underset{
\underset{k_{\Gamma_s} {\rm \ and\ } k_{\bar \Gamma_{1,\ldots,L}}}{{\rm dependent \  on \  } }
}{\underbrace{\det {\bf A}_{\Gamma_s}} }\biggr) \times \underset{
\underset{ k_{\bar \Gamma_{1,\ldots,L}}}{{\rm dependent\  on \  } }
}{\underbrace{\det {\bf A}_{\bar \Gamma_{1,\ldots,L}}} }, \label{factorize_L}
\eal
where  $k_{\Gamma'}$ denotes the set of parameters associated with reactions inside a subnetwork $\Gamma'$. 
The parameter dependence is determined  by using \eqref{factorize'} for every buffering structure $\Gamma_{s}$ ($s=1,\ldots,L$).
For example, \eqref{factorize'} for \red{the buffering structure} $\Gamma_{1}$  implies that the determinant factors except for $\detA_{\Gamma_1}$ is independent of $k_{\Gamma_1}$ and so on.

We can also factorize  a nested sequence of  buffering structures.
Consider a network  $\Gamma$ consisting of a nested sequence of $L$ buffering structures
$\Gamma_1\subset \ldots \subset \Gamma_L:=\Gamma$.
In this case,
the matrix {\bf A} can be written as
\bal
{\bf A} =
\left(\begin{array}{ccccc}
\cellcolor[gray]{.50} {\bf A}_{\Gamma_1} &\cellcolor[gray]{.7} {\mbf *} &\cellcolor[gray]{.85 }  {\mbf *} &\cellcolor[gray]{.95 } \cdots  &{\mbf *}  \\
  \cellcolor[gray]{.7} {\mbf 0} &\cellcolor[gray]{.7} {\bf A}_{\Gamma_2\backslash \Gamma_1}&\cellcolor[gray]{.85 } &\cellcolor[gray]{.95} \cdots &{\mbf *}  \\
\cellcolor[gray]{.85 }     {\mbf 0}  & \cellcolor[gray]{.85 }   {\mbf 0}  &\cellcolor[gray]{.85 }  {\bf A}_{\Gamma_3\backslash \Gamma_2} &\cellcolor[gray]{.95 }& \vdots \\
\cellcolor[gray]{.95}  \vdots  & \cellcolor[gray]{.95} \vdots  &\cellcolor[gray]{.95}  &\cellcolor[gray]{.95} \ddots &  {\mbf *}\\
{\mbf 0}& {\mbf 0} & \cdots &  {\mbf 0} &{\bf A}_{\Gamma_L\backslash \Gamma_{L-1}}
 \end{array}\right),
\eal
where  ${\mbf *}$'s indicate  nonzero blocks and the gradation of shade indicates the nested sequence of buffering structures, $\Gamma_1, \ldots,\Gamma_L$.
By using \eqref{factorize'} iteratively, we can prove
\bal
\det {\bf A}=\underset{
\underset{k_{\Gamma_L}}{{\rm dependent\  on \  } }
}{\underbrace{ \det {\bf A}_{\Gamma_1} }}
\prod_{s=1}^{L-1}
\underset{
\underset{k_{
{\Gamma_{s+1} \backslash {\Gamma_s}}
}
}{{\rm dependent \  on \  } }
}{\underbrace{ \det {\bf A}_{{\Gamma_{s+1}} \backslash {\Gamma_s} } }}\label{factorize_nest}
.
\eal

\begin{figure}[t]
\center
\includegraphics[clip,width=6cm]{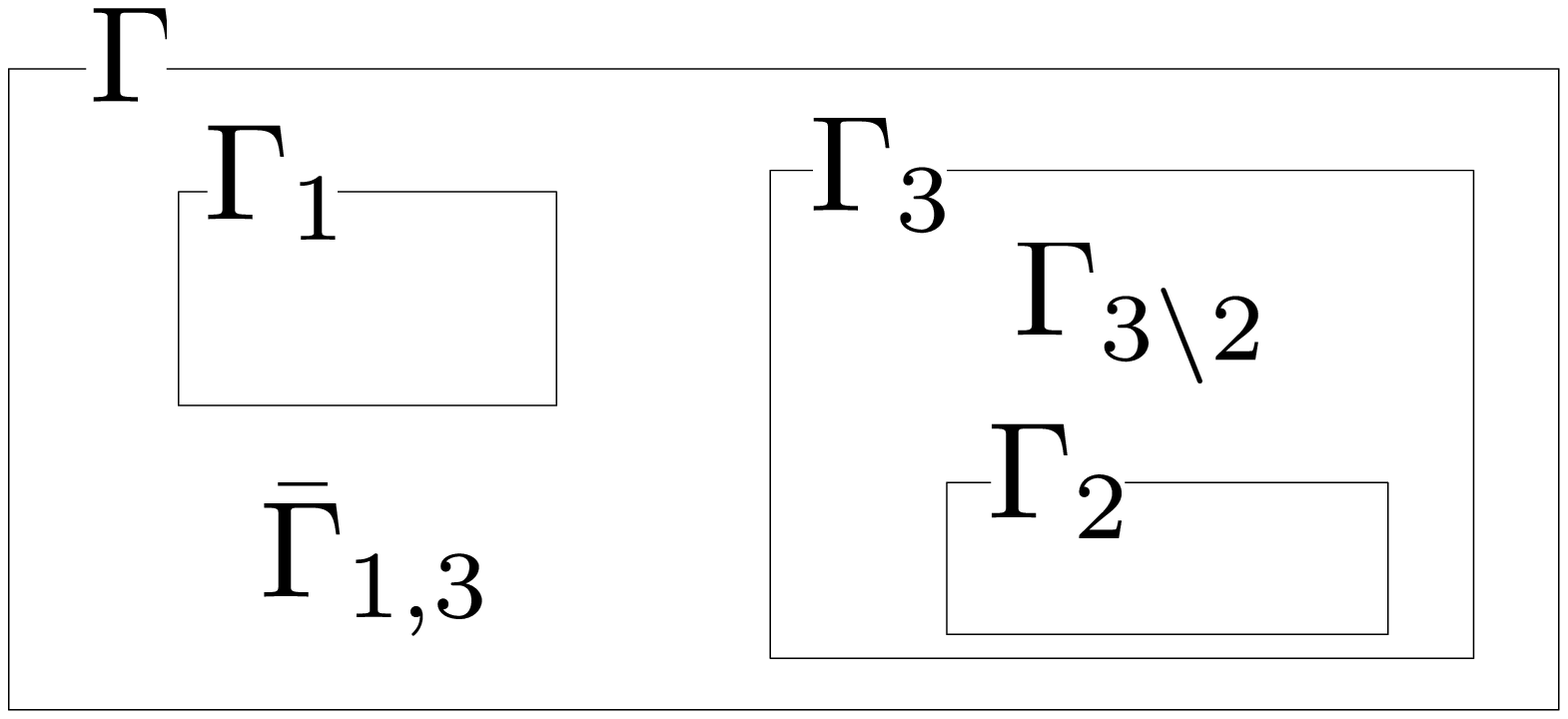}
\caption{
Example network $\Gamma$. Rectangular boxes, $\Gamma_1,\, \Gamma_2,\, \Gamma_3$, indicate buffering structures.
The law of localization indicates that  $k_{\Gamma_1}$ influences  (steady-state values of chemicals and fluxes in) $\Gamma_1$,
$k_{\Gamma_3}$ influences $\Gamma_3$, \red{$k_{\Gamma_{3\backslash 2}}$} influences $\Gamma_2$,
and \red{$k_{\bar \Gamma_{1,3} }$} influences the whole network $\Gamma$.
  }
\label{fig:nw1}
\end{figure}

By using \eqref{factorize_L} and \eqref{factorize_nest}, we can factorize the determinant of {\bf A} for multiple buffering structures and determine  the parameter dependence of each factor. 
We illustrate the procedure of factorization in an example network (see Fig. \ref{fig:nw1}): Suppose that  a whole network $\Gamma $ contains two non-intersecting buffering structures $\Gamma_1$ and $\Gamma_3$, and $\Gamma_3$ further contains a smaller buffering structure $\Gamma_2$ inside it;
\bal
\Gamma& = \Gamma_1 \cup \Gamma_3 \cup\bar \Gamma_{1,3}\non \\
& =  \Gamma_1 \cup(\Gamma_2 \cup \Gamma_{3\backslash 2})\cup\bar \Gamma_{1,3}.
\eal
Here, $
\bar \Gamma_{1,3}:=\Gamma\backslash (\Gamma_1 \cup\Gamma_3)$ and $ \Gamma_{3\backslash 2} \red{:=} \Gamma_3 \backslash \Gamma_2$.
 In this case,
the matrix {\bf A} can be written as
\bal
{\bf A} = \left(\begin{array}{ccc}
\cellcolor[gray]{.8}  {\bf A}_{\Gamma_1} & {\mbf 0} &{\mbf *}  \\
  {\mbf 0} &\cellcolor[gray]{.7}{\bf A}_{\Gamma_3}  & {\mbf *}  \\
 {\mbf 0} & {\mbf 0} & {\bf A}_{\bar \Gamma_{1,3}} \\
 \end{array}\right)
  =
\left(\begin{array}{cccc}
\cellcolor[gray]{.8}  {\bf A}_{\Gamma_1} & {\mbf 0} & {\mbf 0} &{\mbf *}  \\
  {\mbf 0} &\cellcolor[gray]{.7} {\bf A}_{\Gamma_2}& \cellcolor[gray]{.7}  {\mbf *} & {\mbf *}  \\
    {\mbf 0}  &  \cellcolor[gray]{.7}  {\mbf 0}  &\cellcolor[gray]{.9}  {\bf A}_{\Gamma_{3\backslash 2}} &  {\mbf *}\\
 {\mbf 0} & {\mbf 0} & {\mbf 0} & {\bf A}_{\bar \Gamma_{1,3}} \\
 \end{array}\right),
\eal
and the determinant of {\bf A} is factorized as
\bal
\detA &=  \underset{
\underset{k_{\Gamma_1} {\rm \ and\ } k_{\bar \Gamma_{1,3}}}{{\rm depends\  on \  } }
}{\underbrace{\det {\bf A}_{\Gamma_1}} }
\times
\underset{
\underset{k_{\Gamma_3} {\rm \ and\ } k_{\bar \Gamma_{1,3}}}{{\rm depends\  on \  } }
}{\underbrace{\det {\bf A}_{\Gamma_3}} }
\times
\underset{
\underset{k_{\bar \Gamma_{1,3}}}{{\rm depends\  on \  } }
}{\underbrace{ \detA_{\bar\Gamma_{1,3}} }}\non\\
&=  \underset{
\underset{k_{\Gamma_1} {\rm \ and\ } k_{\bar \Gamma_{1,3}}}{{\rm depends\  on \  } }
}{\underbrace{\det {\bf A}_{\Gamma_1}} }
\times
(\underset{
\underset{k_{\Gamma_2}, k_{\Gamma_{3\backslash 2} }, {\rm \ and\ } k_{\bar \Gamma_{1,3}}}{{\rm depends\  on \  } }
}{\underbrace{\det {\bf A}_{\Gamma_2}} }\times
\underset{
\underset{k_{\Gamma_{3\backslash 2}} {\rm \ and\ } k_{\bar \Gamma_{1,3}}}{{\rm depends\  on \  } }
}{\underbrace{\det {\bf A}_{\Gamma_{3\backslash2}}} }
)
\times
\underset{
\underset{k_{\bar \Gamma_{1,3}}}{{\rm depends\  on \  } }
}{\underbrace{ \detA_{\bar\Gamma_{1,3}} }}.\label{factorization_ex}
\eal
In the first line, we have used \eqref{factorize_L} with $L=2$ for the whole network $\Gamma$. In the second line, we have used \eqref{factorize_nest} with $L=2$, namely \eqref{factorize'}, for the buffering structure $\Gamma_3$, where the factor $\detA_{\Gamma_3}$ is factorized further and its dependence on $k_{\Gamma_3}$ is determined more finely.

%

\subsection{ Structural Factorization when  $K_c >0$}\label{sec:fac}

\subsubsection{Buffering structure and the law of localization \red{when} $K_c >0$ (review of \cite{OMpre}) }

The law of localization also holds when $K_c >0$ if we modify the definition of the index $\chi$ appropriately.
Note that the construction of a subnetwork is the same as in the case  $K_c =0$.

 For a subnetwork $\Gamma_s=(\mathfrak m, \mathfrak n)$, we define  $\chi$  as
\bal
\chi (\Gamma_s)=|\mathfrak m| -| \mathfrak n| + (\# {\rm cycle})  - (\# {\rm conserved \ concentration}).\label{chi2}
\eal
Here, $(\# {\rm conserved \ concentration})$ counts conserved quantities in $\Gamma_s$, and is, more precisely, defined as the dimension of the vector space
\bal
V(\mathfrak m) := {\rm span}\, \bigl\{  P^{\mathfrak m}  {\bf u}   |{\bf u} \in\bR^{M},  \, {\bf u}^T \nu ={\bf 0}   \bigr\}, \label{vm}
\eal
where $ P^{\mathfrak m} $ is an $M\times M$ projection matrix for chemicals \red{in} $\mathfrak m$;
\bal
P^{\mathfrak m}_{m,m'} = \delta_{m,m'} {\ \rm if \ } m,m' \in \mathfrak m,\  {\rm otherwise} \ P^{\mathfrak m}_{m,m'}=0.
\eal
Then, we again call a subnetwork $\Gamma_s$ a buffering structure when it satisfies $\chi(\Gamma_s) = 0$, where $\chi$ is defined by \eqref{chi2}.

Recall that, when  $K_c >0$,  steady states generally depend on reaction rate parameters $k_n$ and the initial concentrations $l_\beta$. In \cite{OMpre}, it was proved  that,
{\it for a buffing structure $\Gamma_s$, the values of the steady state concentrations and fluxes outside of $\Gamma_s$ are independent of the reaction rate parameters and conserved quantities (initial conditions) \footnote{We say that a conserved concentraiton $l_\beta$  is associated with $\Gamma_s$, if it can be written as $l_\beta = {\mbf d}_\beta \cdot {\mbf x}$ with $\bf d_\beta \in V(\mathfrak m)$. } associated with $\Gamma_s$.}
This is the generalized version of the law of localization into the case of  $K_c >0$.

\subsubsection{Factorization \red{of the matrix $A$ when $K_c > 0$} }
As discussed in \cite{OMpre}, when $K_c >0$, by taking  appropriate bases for the kernel and cokernel spaces of $\nu$ and permutating indices of the matrix, the matrix ${\bf A}$ can be again expressed as
\bal
 {\bf A} =\begin{pmatrix}
 {\bf A}_{\Gamma_s} & {\bf A}_{\Gamma_s,\bar \Gamma_s} \\
 {\bf 0}_{|\mathfrak m^c| \times |\mathfrak n^c|} & {\bf A}_{\bar \Gamma_s}
 \end{pmatrix},\label{A_coker}
\eal
 where  ${\bf A}_{\Gamma_s}$ is a square matrix, whose column(\red{resp.} row) indices correspond to chemicals and cycles
 (\red{resp.} reactions and conserved concentrations) associated with $\Gamma_s$.

Then, we obtain the following results.
\begin{theorem}
The following hold for  a buffing structure $\Gamma_s$.
\begin{enumerate}
\item[\rm(i)]
The determinant of the  matrix $\bf A$ can be factorized as follows:
\bal
\det {\bf A} = \det {\bf A}_{\Gamma_s} \times \det {\bf A}_{\bar \Gamma_s}. \label{factorize2}
\eal
Thus ${\bf A}_{\Gamma_s, \bar \Gamma_s} $ does not contribute to  $\det {\bf A}$.
\item[\rm(ii)]
\bal
\frac{\p {\bf A}_{\bar \Gamma_s}}{\p k_n}=0, \ \frac{\p {\bf A}_{\bar \Gamma_s}}{\p l_\beta}=0  \  \label{detAind2}
\eal
for reaction rate parameters $k_n$ and conserved concentrations $l_\beta$ associated with $ \Gamma_s$.
\end{enumerate}
\end{theorem}

(ii) can be proved in the same way as in \red{\eqref{detAind}:}
$ {\bf A}_{\bar \Gamma_s}$ is a function of chemical concentrations in $\bar\Gamma_s$, whose steady-state values are independent of reaction rate parameters and conserved concentrations associated with $ \Gamma_s$, due to the law of localization.

%

\subsubsection{\red{Factorization for multiple buffering structures when $K_c  >0$}}

The factorization for multiple buffering structures is  performed in the same way as in the case with $K_c=0$, except that, in the case of $K_c>0$,  determinant factors in $\detA$ depend not only on  $k_n$ but also on that of $l_\beta$. \corr{From the law of localization, we can determine 
\green{how} parameters $k_n$ and conserved concentrations $l_\beta$ influences determinant factors, \green{$\det {\bf A}_{\Gamma_s}$ and $\det {\bf A}_{\bar \Gamma_s}$},  in \eqref{factorize2} structurally.}
\section{Null vectors of  the matrix ${\bf A}$ and  the Jacobian matrix, and bifurcating chemicals}

Suppose that  a chemical reaction system $\Gamma$  exhibits  steady-state bifurcations.
At a bifurcation point,  there exist null vectors\footnote{For  $J$ and $\bf A$, we call their eigenvectors with eigenvalue 0 null vectors, rather than kernel vectors, in order to distinguish them from  kernel vectors of $\nu$. } for the matrix {\rm A} and the Jacobian matrix, since $\detA = {\rm det}\, J =0 $.
In the presence of a buffering structure $\Gamma_s$,  the bifurcation is associated with either $\detA_\gs=0$ or $\detA_\gbs=0$.

In this section,  we show that, for a bifurcation  associated with $\detA_\gs=0$, the null vector ${\mbf v}$ of the Jacobian \red{matrix}
satisfies
\bal
v_m =0  \ {\rm  \ for }\  {\rm chemical \ }  m \in {\bar \Gamma}_s. \label{vm=0}
\eal
Namely,  ${\mbf v}$  has support inside  chemicals in $\Gamma_s$. This implies that, for a bifurcation  associated with $\detA_{\Gamma_s} =0$, only chemicals  inside $\Gamma_s$ undergo a bifurcating behavior.

\subsection{Null vectors of the matrix {\bf A} and the Jacobian matrix when $K_c =0$}

\red{The strategy for showing \eqref{vm=0} is to use the null vectors of ${\rm A}_\gs$ to construct
the associated null vector ${\mbf v}$ of the Jacobian matrix $J$.}

We consider a chemical reaction system  of $M$ chemicals and $N$ reactions
\red{where the cokernel space of $\nu$ consists only of the zero vector.}
Suppose that $\detA_{\Gamma_s}=0$
\red{whose associated buffering structure $\Gamma_s=(\mathfrak m, \mathfrak n)$ contains $|\mathfrak m|$ chemicals,
$|\mathfrak n|$ reactions, and $K_s =  |\mathfrak n|-|\mathfrak m| $ kernel vectors}, $\{ {\bf c}^1,\ldots , {\bf c}^{K_s}\} $, of $\nu$. By construction,  each  \red{${\bf c}^\alpha\in\mathbb{R}^N$} ($\alpha=1,\ldots,K_s$) has support on reactions inside $\Gamma_s$,  and thus  has the following form,
\bal 
{\bf c}^\alpha =
\left(\begin{array}{c}
{\bf c}'^\alpha\\ \hline
{\bf 0}
 \end{array}\right), \label{cc'}
\eal
where the upper $|\mathfrak n|$ components  (\red{resp.} lower $N-|\mathfrak n|$ components) are associated with reactions in $\gs$ (\red{resp.} $\gbs$).
By using ${\bf c}'^\alpha$, ${\bf A}_\gs$ in \eqref{Amat_coker0} can be written as
\bal
 {\bf A}_\gs =\left(\begin{array}{c|cc}
(\dfrac{\p \br  }{\p {\mbf x}})_{\gs} &  -{\bf c}'^1\ldots &  -{\bf c}'^{K_s}
 \end{array}\right),\label{Ags}
\eal
where $(\dfrac{\p \br  }{\p {\mbf x}})_{\gs}$ is an $|\mathfrak n|\times |\mathfrak m|$ matrix whose $(n,m)$ component is given by $\frac{\partial r_n}{\partial x_m}$ \red{with  $n \in \mathfrak n$ and $m \in \mathfrak m$.}   

Since $\detA_{\Gamma_s}=0$, there exist a null vector $\mbf u_s  \in \mathbb{R}^{|\mathfrak m|+K_s}$ such that
\bal
{\bf A}_{\Gamma_s}{\mbf u_s}=0. \label{Aus}
\eal
 If we write $\mbf u_s  =(\eta_1,\ldots, \eta_{|\mathfrak m|}, \zeta_1,\ldots, \zeta_{K_s})$ and substitute ${\bf A}_\gs$ in \eqref{Ags}
 into   \red{Eq.~\eqref{Aus}}, we obtain
 \bal
\sum_{\red{m\in \mathfrak m}} \frac{\partial r_n }{\partial x_m} \eta_m - \sum_{\alpha =1}^{K_s} ({\mbf c'}^{\alpha})_n \zeta_\alpha =0 \label{Ags2}
 \eal
for any reaction \red{$n\in \mathfrak n$.}  Note that all indices in \eqref{Ags2} are associated with $\Gamma_s$.

In order to relate \eqref{Ags2} with the Jacobian \red{matrix} $J$ of the whole system $\Gamma$, we  rewrite \eqref{Ags2} using  indices of the whole system.
First, by using \eqref{cc'} and $\frac{\partial r_n}{\partial x_m} =0$ for \red{$m\in \mathfrak m, n\in \mathfrak n^c$,}  corresponding to the fact that  chemical concentrations in a buffering structure does not appear in  the arguments of   rate functions of reactions outside the buffering structure,
we can \red{rewrite} \eqref{Ags2} as
 \bal
\sum_{\red{m\in \mathfrak m}} \frac{\partial r_n }{\partial x_m} \eta_m - \sum_{\alpha =1}^{K_s} ({\mbf c}^{\alpha})_n \zeta_\alpha =0, \label{Ags3}
 \eal
where $n$ is any reaction in the \red{whole system $\Gamma = (\mathbb{X}, \mathbb{E})$.}  Note that  ${\mbf c}'$ has been replaced  by ${\mbf c}$.
Furthermore, by introducing an $M$-dimensional vector ${\mbf v}$ as ${\mbf v} = ({\mbf \eta},\underset{M-|\mathfrak m| }{\underbrace{0,\ldots, 0}})$,  \eqref{Ags3} can be \red{rewritten} as
  \bal
\sum_{\red{m\in \mathbb{X}}} \frac{\partial r_n }{\partial x_m} \eta_m - \sum_{\alpha =1}^{K_s} ({\mbf c}^{\alpha})_n \zeta_\alpha =0, \label{Ags4}
 \eal
where the summation of $m$ is taken over all chemicals in $\Gamma$.
Finally, by multiplying \eqref{Ags4} with the stoichiometry matrix $\nu$
\red{and using the fact that ${\mbf c}^\alpha$ is a kernel vector of $\nu$}, we obtain
 \bal
\sum_{\red{n\in \mathbb{E}}} \nu_{m',n }\sum_{\red{m\in \mathbb{X}}} \frac{\partial r_n }{\partial x_m} v_{m}=0,
 \eal
or, equivalently,
\bal
\sum_{\red{m\in \mathbb{X}}} J_{m',m}v_m=0.
\eal
Thus, we have proved that $\detA_{\Gamma_s} =0$ is associated with the null vector $\mbf v$ of the Jacobian matrix $J$, whose support is inside chemicals in $\Gamma_s$. Thus, \eqref{vm=0} is proved.

A similar argument cannot be applied into a bifurcation associated with $\detA_\gbs=0$. This difference comes from  the  nonsymmetric structure  of the matrix {\bf A} in \eqref{Amat_coker0}; while columns associated with chemicals and kernels in $\gs$  do not have support on reactions in $\gbs$ (see zero entries in the lower-left block in  \eqref{Amat_coker0}), those associated with $\gbs$  generally have support on both reactions in $\gbs$ and those in $\gs$  (see nonzero entries in the upper-right block in  \eqref{Amat_coker0}). This implies  that, while null vectors of $J$ associated with $\detA_{\gs} =0$ do not have support on  chemicals in $\gbs$,  those associated with $\detA_{\gbs} =0$ generally have support on both chemicals in $\gbs$ and those in $\gs$.

\subsection{Null vectors of the matrix {\bf A} and the Jacobian matrix when  $K_c >0$}

A similar result can also be proved when $K_c >0$.
Let $\Gamma$ be a network system of $M$ chemicals and $N$ reactions
whose corresponding stoichiometry matrix $\nu$ has $K$ kernel vectors and $K_c>0$ cokernel vectors (conserved concentrations).

\corr{Suppose that the subnetwork  $\gs = (\mathfrak m, \mathfrak n)$ is a buffering structure.  We separate the set of all chemicals $\mathbb{X}$ into two groups, $\mathbb{X}_{dep}$ and $\mathbb{X}_{ind}$, as in section \ref{sec:eq>0}. Note that $K_c = |\mathbb{X}_{dep}|$ since each of cokernel vectors of $\nu$  corresponds to a chemical in $\mathbb{X}_{dep}$. \footnote{More explicitly, this correspondence is expressed by the identity matrix ${\bf 1}_{K_c}$ in \eqref{Amat2}, where the columns are associated with $\mathbb{X}_{dep}$ and the rows with the cokernel vectors.} Accordingly, we also separate the set of chemicals $\mathfrak m$  in $\gs$ into two groups, $\mathfrak m_{dep}:=\mathfrak m \cup  \mathbb{X}_{dep}$ and $\mathfrak m_{ind}:=\mathfrak m \cup  \mathbb{X}_{ind}$.  Then, this buffering structure consists of $|\mathfrak m|= |\mathfrak m_{dep}| + | \mathfrak m_{ind}|$ chemicals, $|\mathfrak n|$ reactions, $K_s$ kernel vectors of $\nu$, and $|\mathfrak m_{dep}|$ conserved concentrations, satisfying $|\mathfrak m| -  |\mathfrak n| + K_s -|\mathfrak m_{dep}|=0 $. }

The matrix ${\bf A}_\gs$ in \eqref{A_coker} has the following form,
\bal
 {\bf A}_\gs =
 \left(\begin{array}{cc|c}
(\dfrac{\p \br  }{\p {\mbf x}_{ind}})_{\gs} & (\dfrac{\p \br  }{\p {\mbf x}_{dep} } )_{\gs} &  -{\bf c}'^1\ldots   -{\bf c}'^{K_s}
\\ \hline
 - L_\gs & {\bf 1}_{K^s_c}  & {\bf 0}_{K^s_c\times K_s}
  \end{array}
 \right).\label{Ags_coker}
\eal
Here, the column indices are associated with $\mathfrak m_{ind}$, $\mathfrak m_{dep}$, and kernel vectors of $\nu$ in $\Gamma_s$, from left to right, and the row indices are associated with $\mathfrak n$ and cokernel vectors associated with  $\gs$. 
The  left two blocks  of ${\bf A}_\gs$ are obtained by taking columns and rows associated with $\Gamma_s$  from the left two block of the matrix {\bf A} in \eqref{Amat2}: 
$(\dfrac{\p \br  }{\p {\mbf x}_{ind}})_{\gs}$ is an $N_s \times |\mathfrak m_{ind}|$ matrix, $(\dfrac{\p \br  }{\p {\mbf x}_{dep}})_{\gs}$ is an $N_s \times| \mathfrak m_{dep}|$ matrix.
$L_\gs$ is an $|\mathfrak m_{dep}| \times  |\mathfrak m_{ind}|$ matrix, which is  the submatrix of the matrix $L$ in \eqref{Amat2}, \corr{  obtained by taking the rows  associated with   $\mathfrak m_{dep}$ and columns associated with $\mathfrak m_{ind}$.}

Suppose that $\detA_\gs=0$. We denote the null vector $\mbf u_s  \in \mathbb{R}^{|\mathfrak m|+K_s} $ of ${\bf A}_\gs$ as
\bal
{\mbf u}_s = (\eta_1, \ldots, \eta_{ |\mathfrak m_{ind}|}, \tilde \eta_{1},\ldots,\tilde\eta_{ |\mathfrak m_{dep}|}, \zeta_1,\ldots, \zeta_{K_s})^T.
\eal
The equation ${\bf A}_\gs{\mbf u}_s =0$ leads to the following two equations;
\bal
\sum_{m \in  \mathfrak m_{ind}} \frac{\partial r_n }{\partial x_{ind,m}} \eta_m +\sum_{m \in  \mathfrak m_{dep}} \frac{\partial r_n }{\partial x_{dep,m}} \tilde\eta_m
- \sum_{\alpha =1}^{K_s} ({\mbf c'}^{\alpha})_n \zeta_\alpha =0 \label{au2}
\eal
for  any reaction $ n \in \mathfrak n$, and
\bal
 \tilde \eta_{m}= \sum_{m' \in  \mathfrak m_{ind}}(L)_{m,m'} \eta_{m'}, \label{tildeeta}
\eal
for  any  chemical $m \in \mathfrak m_{dep}$.

By plugging \eqref{tildeeta} into \eqref{au2}, we obtain
\bal
\sum_{m \in  \mathfrak m_{ind}} (\frac{\partial r_n }{\partial x_{ind,m}}  +\sum_{m'\in  \mathfrak m_{dep}} \frac{\partial r_n }{\partial x_{dep,m'}}L_{m',m})\eta_m- \sum_{\alpha =1}^{K_s} ({\mbf c'}^{\alpha})_n \zeta_\alpha =0, \label{au3}
\eal
for  any reaction $ n \in \mathfrak n$. 

As in the case of $K_c =0$, we rewire \eqref{au3} using indices of the whole systems.
We introduce an $|\mathbb{X}_{ind}| (= M-K_c)$ dimensional vector,
\bal
{\mbf v} = ({\mbf \eta}, \underset{\tilde M  -| \mathfrak m_{ind}|}{\underbrace{0,\ldots, 0}}) \in \mathbb{R}^{\tilde M }.
\eal
Then, by using the same reasoning used to derive \eqref{Ags4} from \eqref{Ags2} and the relation $L_{m'm} =0$ for $m'\in \mathbb{X}_{dep}\backslash \mathfrak m_{dep}, m\in \mathfrak m_{ind}$, \footnote{
This can be proved from the condition of a buffering structure: If $L_{m'm} $ was nonzero for some $m'\in \mathbb{X}_{dep}\backslash \mathfrak m_{dep}, m\in \mathfrak m_{ind}$,  nonzero entries would appear in the lower-right block of {\bf A} in \eqref{A_coker} and $\gs$ would not be a buffering structure, which contradicts our assumption.   }    \eqref{au3} can be rewritten as 
\bal
\sum_{m \in  \mathbb{X}_{ind}} (\frac{\partial r_n }{\partial x_{ind,m}}  +\sum_{m' \in   \mathbb{X}_{dep}} \frac{\partial r_n }{\partial x_{dep,m'}}L_{m',m})v_m- \sum_{\alpha =1}^{K_s} ({\mbf c}^{\alpha})_n \zeta_\alpha =0, \label{au4}
\eal
where we have replaced $N_s$-dimensional vectors ${\mbf c}'^{\alpha}$ by $N$-dimensional vectors ${\mbf c}^{\alpha}$ (see \eqref{cc'}), and  the summations over $m,m'$ are taken over all chemicals in $\Gamma$.

Finally, by multiplying \eqref{au4} with $\tilde\nu$ , we obtain
\bal
\sum_{n\in \mathbb{E}} {\tilde \nu}_{m'',n}\sum_{m\in \mathbb{X}_{ind}} (\frac{\partial r_n }{\partial x_{ind,m}}  +\sum_{m'\in  \mathbb{X}_{dep}} \frac{\partial r_n }{\partial x_{dep,m'}}L_{m',m})v_m=0,
\eal
or, equivalently,
\bal
\sum_{m\in \mathbb{X}_{ind}} \tilde J_{m', m}v_m=0,
\eal
where $\tilde J$ is defined in \eqref{J'0}. Thus, the null vector $\mbf v$ of $\tJ$ associated with $\detA_\gs=0$ has support in chemicals in $\gs$.

\section{Parameters used in Fig. 3  in the main text}
\corr{The dynamics of the hypothetical example in the main text is given  by the following ODEs,
\bal
\dot x_A &= k_1 - k_2 x_{\text{A}}+k_3  x_{\text{B}}\left(1+ \frac{k_{3,\text{A}} x_{\text{A}}^2}{x_{\text{A}}^2+5}\right)\non\\
\dot x_B &=  k_2 x_{\text{A}} -k_3  x_{\text{B}}\left(1+ \frac{k_{3,\text{A}} x_{\text{A}}^2}{x_{\text{A}}^2+5}\right) -k_4 x_{\text{B}} \left(1+ \frac{k_{4,\rm{C}} x_{\text{C}}^2}{x_{\text{C}}^2+5}\right) \non\\
\dot x_C &= k_4 x_{\text{B}} \left(1+ \frac{k_{4,\rm{C}} x_{\text{C}}^2}{x_{\text{C}}^2+5}\right) - k_5 x_{\text{C}}\left(1+ \frac{k_{5,\rm{B}} x_{\text{B}}^2}{x_{\text{B}}^2+5}\right) .
\eal
}
\corr{In Fig. 3 in the main text, we used the following parameters.}
For the bifurcation associated with  ${\bar{\Gamma}_s}$ (Fig. 3 (a)), we used the following parameter sets.
For the plot of \red{the curve of} red squares, $\vec k=(80, \underline{81},\underline{ 25}, 46, 43) $, $(k_{3,{\rm A}},
k_{4,{\rm C}})=(\underline{1},70)$, \red{while for}
the plot of \red{the curve of} blue circles, $\vec k=(80,\underline{ 85},\underline{ 20}, 46, 43)$ \red{and}  $(k_{3,{\rm A} }, k_{4,{\rm C}})=(\underline{5},70)$.
\red{Note that we use the underlines to emphasize which components in the parameter sets are different in these two considered cases.}

For the bifurcation associated with  ${\Gamma_s}$ (Fig. 3 (b)), we used the following parameter sets.
For the plot of \red{the curve of} red squares, $\vec k =(k_1,\ldots,k_5)=(7, 54, 8, \underline{22},\underline{ 45}) $, $(k_{4,{\rm C}}, k_{5,{\rm B}})=(\underline{0},\underline{0})$,
\red{while for the plot of the curve of} blue circles, $\vec k=(7, 54, 8, \underline{25}, \underline{40})$ \red{and} $(k_{4,{\rm C}}, k_{5,{\rm B}})=(\underline{5},\underline{5})$.
\red{As before, we use the underlines to emphasize the different components between these two considered parameter sets.}

For the bifurcation associated with  ${\bar{\Gamma}_s}$ (Fig. 3 (c)), we used the following parameter sets.
For the plot of \red{the curve of} red squares, $\vec k=(7, 54,8 , 25, 40) $, $(k_{3,{\rm A}}, k_{4,{\rm C}})=(\underline{100},5)$,
\red{while for} the plot of \red{the curve of} blue circles, $\vec k=(7, 54,8 , 25, 40)$,  $(k_{3,{\rm A} }, k_{4,{\rm C}})=(\underline{99.9},5)$.
\red{The use of the underlines can be explained as before.}

\section{{Structural bifurcation analysis for the E. coli network}}

\subsection{Reaction list  of the E. coli network}
The central carbon metabolism  of the E. coli  in the  main text consists of the following reactions:

{\small 
1: Glucose  +  PEP  $\rightarrow$  G6P  +  PYR. 
 
 2: G6P   $\leftarrow$  F6P. 
 
 3: F6P   $\rightarrow$  G6P.
  
  4: F6P  $\rightarrow$  F1,6P. 
   
   5: F1,6P  $\rightarrow$  G3P  +  DHAP. 
   
   6: DHAP   $\rightarrow$  G3P.
   
 7: G3P   $\rightarrow$  3PG. 
  
  8: 3PG   $\rightarrow$  PEP. 
  
  9: PEP   $\rightarrow$  3PG.  
  
  10: PEP   $\rightarrow$  PYR. 
  
  11: PYR   $\rightarrow$  PEP. 
  
  12: PYR    $\rightarrow$  AcCoA  +   CO2. 
  
  13: G6P  $\rightarrow$  6PG. 
  
  14: 6PG  $\rightarrow$   Ru5P  +  CO2. 
  
  15: Ru5P  $\rightarrow$  X5P.
  
   16: Ru5P  $\rightarrow$   R5P. 
   
   17: X5P  +  R5P  $\rightarrow$   G3P  +  S7P. 
   
   18: G3P  +  S7P  $\rightarrow$   X5P  +  R5P. 
   
   19: G3P  +  S7P  $\rightarrow$   F6P  +  E4P.
   
    20: F6P  +  E4P  $\rightarrow$  G3P  +  S7P. 
    
21:  X5P  +  E4P  $\rightarrow$   F6P  +  G3P. 
  
  22: F6P   +  G3P  $\rightarrow$   X5P  +  E4P.  
  
  23: AcCoA  +    $\rightarrow$  CIT. 
  
  24: CIT   $\rightarrow$  ICT. 
  
  25: ICT  $\rightarrow$  2${\rm \mathchar`-}$KG  +  CO2. 
  
  26: 2-KG  $\rightarrow$   SUC  +  CO2.
  
   27: SUC   $\rightarrow$  FUM. 
   
  28:  FUM  $\rightarrow$  MAL. 
   
   29: MAL   $\rightarrow$  OAA.
   
 30: OAA   $\rightarrow$  MAL.

 31: PEP  +  CO2  $\rightarrow$  OAA.

 32: OAA  $\rightarrow$  PEP  +   CO2. 

 33: MAL  $\rightarrow$   PYR  +  CO2.

34: ICT   $\rightarrow$  SUC  +  Glyoxylate. 

 35: Glyoxylate  +  AcCoA  $\rightarrow$  MAL. 

 36: 6PG  $\rightarrow$   G3P  +  PYR. 

 37: AcCoA  $\rightarrow$   Acetate. 

38:  PYR  $\rightarrow$  Lactate. 

 39: AcCoA  $\rightarrow$  Ethanol. 

 40: R5P  $\rightarrow$ (output).

 41: OAA  $\rightarrow$ (output).

 42: CO2  $\rightarrow$ (output).

43:  (input) $\rightarrow$  Glucose. 

 44:  Acetate $\rightarrow$ (output).
 
  45: Lactate $\rightarrow$ (output).

46:  Ethanol $\rightarrow$ (output).
}

\subsection{Buffering structures in the E. coli network}\label{sec:bsec}
\begin{figure}[h]
\center
\includegraphics[clip,width=14cm]{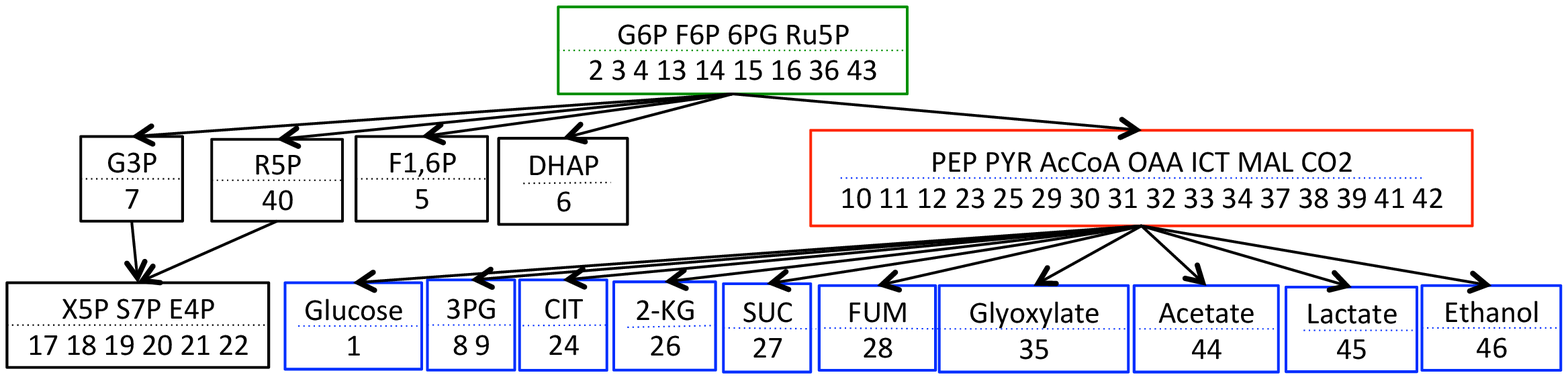}
\caption{The set of chemicals and reactions, (indicated by numbers) in each box corresponds to a buffering structure with subtraction of its inner buffering structures. 
  }
\label{fig:ecbs}
\end{figure}

Assuming that each reaction rate function depends on its substrate concentrations, we find 17 independent\footnote{In general, the union of buffering structures is also a buffering structure. Therefore, precisely speaking, these 17 buffering structures should be regarded as ``generators'' of  buffering structures, since  we can construct all buffering structures in the system by taking all possible unions of these 17 buffering structures. }  buffering structures in the E. coli system \cite{OM}. The inclusion relation between them is summarized in Fig. \ref{fig:ecbs}. For each box in Fig. \ref{fig:ecbs},  the set of chemicals and reactions (indicated by numbers) in the box plus those in its downward boxes gives a buffering structure. In other words, the set of chemicals and reactions in each box gives a subnetwork, which is a buffering structure with subtraction of its inner buffering structures.  
\footnote{Remark that,  in Fig. \ref{fig:ecbs},  while a box without emanating arrows from it, such as $( \{\rm Glucose\} ,\{1 \})$, corresponds to a buffering structure, a box with emanating arrows from it, such as $( \{\rm R5P\} ,\{40 \})$,  itself is  not a buffering structure. } The determinant $\detA$ are factorized according to these 17 subnetworks. 

We write down explicitly the 17 buffering structures:

{\small 
$\Gamma_1=(\{ \rm 
Glucose
\},\{
1
\})$.

$\Gamma_2=(\{ \rm 
3PG
\},\{
8 
\})$

$\Gamma_3=(\{ \rm 
CIT
\},\{
24 
\})$,

$\Gamma_{4}=(\{ \rm 
2{\rm \mathchar`-KG}
\},\{
26 
\})$,

$\Gamma_{5}=(\{ \rm 
SUC
\},\{
27 
\})$,

$\Gamma_{6}=(\{ \rm 
FUM
\},\{
28 
\})$,

$\Gamma_{7}=(\{ \rm 
Glyoxylate
\},\{
35 
\})$,

$\Gamma_{8}=(\{ \rm 
Acetate
\},\{
44 
\})$,

$\Gamma_{9}=(\{ \rm 
Lactate
\},\{
45 
\})$,

$\Gamma_{10}=(\{ \rm 
Ethanol
\},\{
46 
\})$,

$\Gamma_{11}=(\{ \rm 
Glucose,
PEP,
3PG,
PYR,
AcCoA,
OAA,
CIT,
ICT,\\
2{\rm \mathchar`-KG},
SUC,
FUM, 
MAL,
CO2,
Glyoxylate,
Acetate,
Lactate,\\
Ethanol
\},\{
1,
8 ,
9 ,
10 ,
11 ,
12 ,
23 ,
24 ,
25 ,
26 ,
27 ,
28 ,
29 ,
30 ,
31 ,\\
32 ,
33 ,
34 ,
35 ,
37 ,
38 ,
39 ,
41 ,
42 ,
44 ,
45 ,
46 
\})$,

$\Gamma_{12}=(\{ \rm 
X5P,
S7P,
E4P
\},\{
17 ,
18 ,
19 ,
20 ,
21 
\})$,

$\Gamma_{13}=(\{ \rm 
G3P,
X5P,
S7P,
E4P
\},\{
7 ,
17 ,
18 ,
19 ,
20 ,
21 ,
22 
\})$

$\Gamma_{14}=(\{ \rm 
X5P,
R5P,
S7P,
E4P
\},\{
17 ,
18 ,
19 ,
20 ,
21 ,
40 
\})$,

$\Gamma_{15}=(\{ \rm 
F1,6P
\},\{
5 
\})$,

$\Gamma_{16}=(\{ \rm 
DHAP
\},\{
6 
\})$,

$\Gamma_{17}=(\{ \rm 
Glucose,
PEP,
G6P,
F6P,
F1,6P,
DHAP,
G3P,
3PG,\\
PYR,
6PG,
Ru5P,
X5P,
R5P, 
S7P,
E4P,
AcCoA,
OAA,
CIT,\\
ICT,
2{\rm \mathchar`-KG},
SUC,
FUM,
MAL,
CO2,
Glyoxylate,
Acetate,\\
Lactate,
Ethanol
\},  \{
1,
2 ,
3 ,
4 ,
5 ,
6 ,
7 ,
8 ,
9 ,
10 ,
11 ,
12 ,
13 ,
14 ,
15 ,\\
16 ,
17 ,
18 ,
19 ,
20 ,
21 ,
22 ,
23 ,
24 ,
25 ,
26 ,
27 , 
28 ,
29 ,
30 ,
31 ,
32 ,
33 ,\\
34 ,
35 ,
36 ,
37 ,
38 ,
39 ,
40 ,
41 ,
42 ,
44 ,
45 ,
46 
\})$.

}
\subsection{Structural bifurcation analysis} 
For the E. coli system, we construct  the matrix {\bf A}. 
Up to a constant factor,\footnote{The overall constant  of $\detA$, which depends on normalization of kernel vectors of $\nu$ in {\bf A}, is irrelevant for our discussion. } the determinant $\detA$ is factorized as  
\bal
\detA =
& r_{1,\text{Glucose}}\, r_{5,\text{F1,6P}}\, r_{6,\text{DHAP}}\, r_{8,\text{3PG}} \,  r_{44,\text{Acetate}}\,  r_{45,\text{Lactate}} \, r_{46,\text{Ethanol}}
\non \\
\times& r_{24,\text{CIT}} \,r_{26,\text{2-KG}}\, r_{27,\text{SUC}}\, r_{28,\text{FUM}} \, r_{35,\text{Glyoxylate}} \non \\
\times& r_{7,\text{G3P}}\, r_{40,\text{R5P}}\, \left(r_{17,\text{X5P}} r_{19,\text{S7P}} r_{21,\text{E4P}}+r_{18,\text{S7P}} r_{20,\text{E4P}} r_{21,\text{X5P}}\right)\non \\
\times & \detA_{\Gamma'}\times \detA_{\Gamma''}\non\\
\label{detAec}
\eal
Here, $r_{n,m} = \frac{\partial r_n}{\partial x_m}|_{\mbf x = \mbf x^*}$. Each of the 17 factors in \eqref{detAec} is associated with a  subnetwork in Fig. \ref{fig:ecbs}.
Here, the first two lines is the product of the factors $\detA_{\Gamma_s}$($s=1,3,4,6,9,10,$ $ 11,12,13,15,16,17$), each of which is associated with a buffering structure  $\Gamma_s$ with a single chemical. In the third line, $r_{7,\text{G3P}}$,  $r_{40,\text{R5P}}$, and the factor inside (...) are associated with the three subnetworks, $(\{{\rm G3P}\},\{7\})$, $(\{{\rm R5P}\},\{40\})$, and $\Gamma_{12} = (\{{\rm X5P,S7P,E4P}\},\{17,18,19,20,21,22\})$ in Fig \ref{fig:ecbs}, respectively.  
The factor $\detA_{\Gamma'}$ in the last line of  \eqref{detAec} is associated with the subnetwork $\Gamma': =  (\{{\rm G6P,F6P,6PG,Ru5P}\},\{2,3,4,13,14,15,16,36,43\})$ (the green box  in Fig. \ref{fig:ecbs}) and  given by
\bal
\detA_{\Gamma'} &= r_{2,\text{G6P}} r_{4,\text{F6P}} \left(r_{15,\text{Ru5P}}+r_{16,\text{Ru5P}}\right) \left(r_{14,\text{6PG}}+r_{36,\text{6PG}}\right)\non\\ 
& +r_{13,\text{G6P}}  \biggl(r_{4,\text{F6P}} \left(r_{15,\text{Ru5P}}+r_{16,\text{Ru5P}}\right) \left(r_{14,\text{6PG}}+r_{36,\text{6PG}}\right)\non\\ 
&+r_{3,\text{F6P}} \bigl(r_{14,\text{6PG}} r_{16,\text{Ru5P}}+\left(r_{15,\text{Ru5P}}+r_{16,\text{Ru5P}}\right) r_{36,\text{6PG}}\bigr)\biggr).
\eal
The last factor  $\detA_{\Gamma''}$ in \eqref{detAec} is associated with the subnetwork $\Gamma'': =  (\{{\rm PEP,PYR,AcCOA,OAA,}$ ${\rm ICT,MAL,CO2}\},$ $\{10,11,12,23,25,29,30,31,32,33,34,37,38,39,41,42\})$ (\green{the red box} in Fig. \ref{fig:ecbs}) and given by
{\footnotesize
\bal
\detA_{\Gamma''} &= r_{10,\text{PEP}} (r_{38,\text{PYR}} (r_{23,\text{OAA}} (r_{37,\text{AcCoA}}+r_{39,\text{AcCoA}}) (r_{25,\text{ICT}} (2 r_{29,\text{MAL}} r_{31,\text{CO2}} \non \\ & +r_{33,\text{MAL}} (2 r_{31,\text{CO2}}-r_{42,\text{CO2}})
+r_{34,\text{ICT}} (r_{33,\text{MAL}} (r_{31,\text{CO2}}-r_{42,\text{CO2}})+r_{29,\text{MAL}} (r_{31,\text{CO2}}+r_{42,\text{CO2}})))
 \non \\ & -(r_{23,\text{AcCoA}} (r_{25,\text{ICT}}+2 r_{34,\text{ICT}})+(r_{25,\text{ICT}}+r_{34,\text{ICT}}) (r_{37,\text{AcCoA}}
+r_{39,\text{AcCoA}})) (r_{29,\text{MAL}} (r_{31,\text{CO2}} r_{41,\text{OAA}}
\non \\ &+(r_{32,\text{OAA}}+r_{41,\text{OAA}}) r_{42,\text{CO2}})+r_{33,\text{MAL}} (r_{31,\text{CO2}} r_{41,\text{OAA}}  +(r_{30,\text{OAA}}+r_{32,\text{OAA}}
+r_{41,\text{OAA}}) r_{42,\text{CO2}}))) 
\non \\ & +r_{12,\text{PYR}} (-(r_{25,\text{ICT}}+r_{34,\text{ICT}}) (r_{37,\text{AcCoA}}+r_{39,\text{AcCoA}}) (r_{29,\text{MAL}} (2 r_{31,\text{CO2}} r_{41,\text{OAA}} \non \\ & +(r_{32,\text{OAA}}+r_{41,\text{OAA}}) r_{42,\text{CO2}})
+r_{33,\text{MAL}} (2 r_{31,\text{CO2}} r_{41,\text{OAA}}  +(r_{30,\text{OAA}}+r_{32,\text{OAA}}+r_{41,\text{OAA}}) r_{42,\text{CO2}}))
 \non \\ & +r_{23,\text{OAA}} (r_{37,\text{AcCoA}}+r_{39,\text{AcCoA}}) (r_{25,\text{ICT}} (2 r_{29,\text{MAL}} r_{31,\text{CO2}}+r_{33,\text{MAL}} (2 r_{31,\text{CO2}}-r_{42,\text{CO2}}))
\non \\ &+r_{34,\text{ICT}} (r_{33,\text{MAL}} (2 r_{31,\text{CO2}}-r_{42,\text{CO2}})+r_{29,\text{MAL}} (2 r_{31,\text{CO2}}+r_{42,\text{CO2}})))
\non \\ &-r_{23,\text{AcCoA}} (r_{25,\text{ICT}} (r_{33,\text{MAL}} (4 r_{31,\text{CO2}} r_{41,\text{OAA}}+(r_{30,\text{OAA}}
\non \\ &+r_{32,\text{OAA}}) r_{42,\text{CO2}})+r_{29,\text{MAL}} (4 r_{31,\text{CO2}} r_{41,\text{OAA}}+(r_{32,\text{OAA}}+r_{41,\text{OAA}}) r_{42,\text{CO2}}))
\non \\ &+r_{34,\text{ICT}} (r_{33,\text{MAL}} (4 r_{31,\text{CO2}} r_{41,\text{OAA}}+(r_{30,\text{OAA}}+r_{32,\text{OAA}}) r_{42,\text{CO2}})
\non \\ &+r_{29,\text{MAL}} (4 r_{31,\text{CO2}} r_{41,\text{OAA}}+(r_{32,\text{OAA}}+2 r_{41,\text{OAA}}) r_{42,\text{CO2}})))))
\non \\ &-r_{31,\text{PEP}} (-r_{11,\text{PYR}} (r_{29,\text{MAL}}+r_{33,\text{MAL}}) (r_{23,\text{OAA}} r_{34,\text{ICT}} (r_{37,\text{AcCoA}}+r_{39,\text{AcCoA}})
\non \\ &-(r_{23,\text{AcCoA}} (r_{25,\text{ICT}}+2 r_{34,\text{ICT}})+(r_{25,\text{ICT}}+r_{34,\text{ICT}}) (r_{37,\text{AcCoA}}+r_{39,\text{AcCoA}})) r_{41,\text{OAA}})
\non \\ &+r_{38,\text{PYR}} (r_{23,\text{AcCoA}} (r_{25,\text{ICT}}+2 r_{34,\text{ICT}}) (r_{30,\text{OAA}} r_{33,\text{MAL}}+(r_{29,\text{MAL}}+r_{33,\text{MAL}}) r_{41,\text{OAA}})
\non \\ &+(r_{37,\text{AcCoA}}+r_{39,\text{AcCoA}}) (r_{23,\text{OAA}} (r_{25,\text{ICT}} r_{33,\text{MAL}}+(r_{33,\text{MAL}}-r_{29,\text{MAL}}) r_{34,\text{ICT}})
\non \\ &+(r_{25,\text{ICT}}+r_{34,\text{ICT}}) (r_{30,\text{OAA}} r_{33,\text{MAL}}+(r_{29,\text{MAL}}+r_{33,\text{MAL}}) r_{41,\text{OAA}})))
\non \\ &+r_{12,\text{PYR}} (r_{23,\text{AcCoA}} (r_{25,\text{ICT}} (r_{30,\text{OAA}} r_{33,\text{MAL}}
\non \\ &+r_{29,\text{MAL}} r_{41,\text{OAA}})+r_{34,\text{ICT}} (r_{30,\text{OAA}} r_{33,\text{MAL}}+2 r_{29,\text{MAL}} r_{41,\text{OAA}}))+(r_{37,\text{AcCoA}}
\non \\ &+r_{39,\text{AcCoA}}) (r_{23,\text{OAA}} (r_{25,\text{ICT}} r_{33,\text{MAL}}+(r_{33,\text{MAL}}-r_{29,\text{MAL}}) r_{34,\text{ICT}})
\non \\ &+(r_{25,\text{ICT}}+r_{34,\text{ICT}}) (r_{30,\text{OAA}} r_{33,\text{MAL}}+(r_{29,\text{MAL}}+r_{33,\text{MAL}}) r_{41,\text{OAA}})))) r_{42,\text{CO2}}. 
\eal
}

Since $r_{n,m}>0$, we see that, among the 17 factors in \eqref{detAec}, only the last factor $\detA_{\Gamma''}$ contains both of plus and minus signs. Therefore,   if this system exhibits a steady-state bifurcation (under parameter change), it should be the subnetwork $\Gamma''$ whose determinant changes its sign at the bifurcation point.

\subsection{Numerical analysis} 
In the discussion so far, we have not assumed any specific kinetic for reaction rates. 
To  numerically demonstrate bifurcation behaviors  in the E. coli system, we first  consider the case that  all the reactions obey the mass-action kinetics with reaction rate constant $k_n$ $(n=1,\ldots,47)$.  In this case, we found that, for any parameter choices, the E. coli system has either a single stable solution or a blow up solution, and  no steady-state bifurcations were observed. We also performed the same analysis in the case of  the Michaelis-Menten kinetics, and no  bifurcations were observed. 

We next consider the case that  the reaction 11 : $\rm PYR \rightarrow \rm PEP$  is positively regulated by PEP. Specifically, we modified the rate of reaction 11 from $r_{11} = k_{11} x_{\rm PYR}$ into 
\bal
r_{11} = k_{11} x_{\rm PYR} \biggl(1 + k_{11,PEP}\frac{x_{\rm PEP}^2}{x_{\rm PEP}^2 +K }\biggr)\label{pepreg}
\eal
where $ k_{11,\rm PEP}$ represents the strength of the regulation.  All reactions except reaction 11 obey the mass-action kinetics as before. 

We remark that the regulation from PEP to reaction 11 does not change the buffering structures in section \ref{sec:bsec}  since adding this regulation does not ruin the condition of output-completeness. \footnote{This is generally the case, if an additional regulation is within  a buffering structure $\gs$, i.e. from a chemical in $\gs$  to a reaction in $\gs$. } Thus,  the inclusion relation of buffering structures shown in Fig. \ref{sec:bsec} is also intact  under the modification \eqref{pepreg} of the kinetics.

As explained previously, only bifurcations associated with $\Gamma''$ (the red box in Fig \ref{fig:ecbs}) are possible for this system. The inducing parameters  are then  given by  parameters associated with  reactions \green{in} the green and red boxes in  Fig \ref{fig:ecbs}. 
As a candidate bifurcation parameter, we \green{choose}  the parameter $ k_{11,PEP}$, which is associated with reaction 11 and an inducing parameter for $\Gamma''$. The reaction rate constants of the mass-action kinetics were set as 
{\small
\bal
(k_1,\ldots, k_{47}) &=(18.9, \, 55.9, \, 20.5, \, 5.17, \, 8.14, \, 107, \, 15.7, \, 2.38, \, 32, \, 3.08, \,
 \non\\& 38.8, \,  471, \, 7.54, \, 1.28, \, 24.4, \, 84.9, \, 23.1, \, 1.64, \, 90.6, \, 132, \, 
 \non\\&65, \, 47, \, 237, \, 1.19, \, 11.7, \,  1.80, \, 98.5, \, 27, \,  1090, \, 3.15, \, 
 \non\\& 484, \, 1.44, \, 543, \, 12.2, \, 20, \,  89.5, \, 2.98, \, 7.23, \,  48.9, \, 2.96, \,
 \non\\& 21.6, \, 37.6, \, 85, \, 131, \, 28.2, \, 2.37).
\eal
}
 \begin{figure}[h]
\center
\includegraphics[clip,width=12cm]{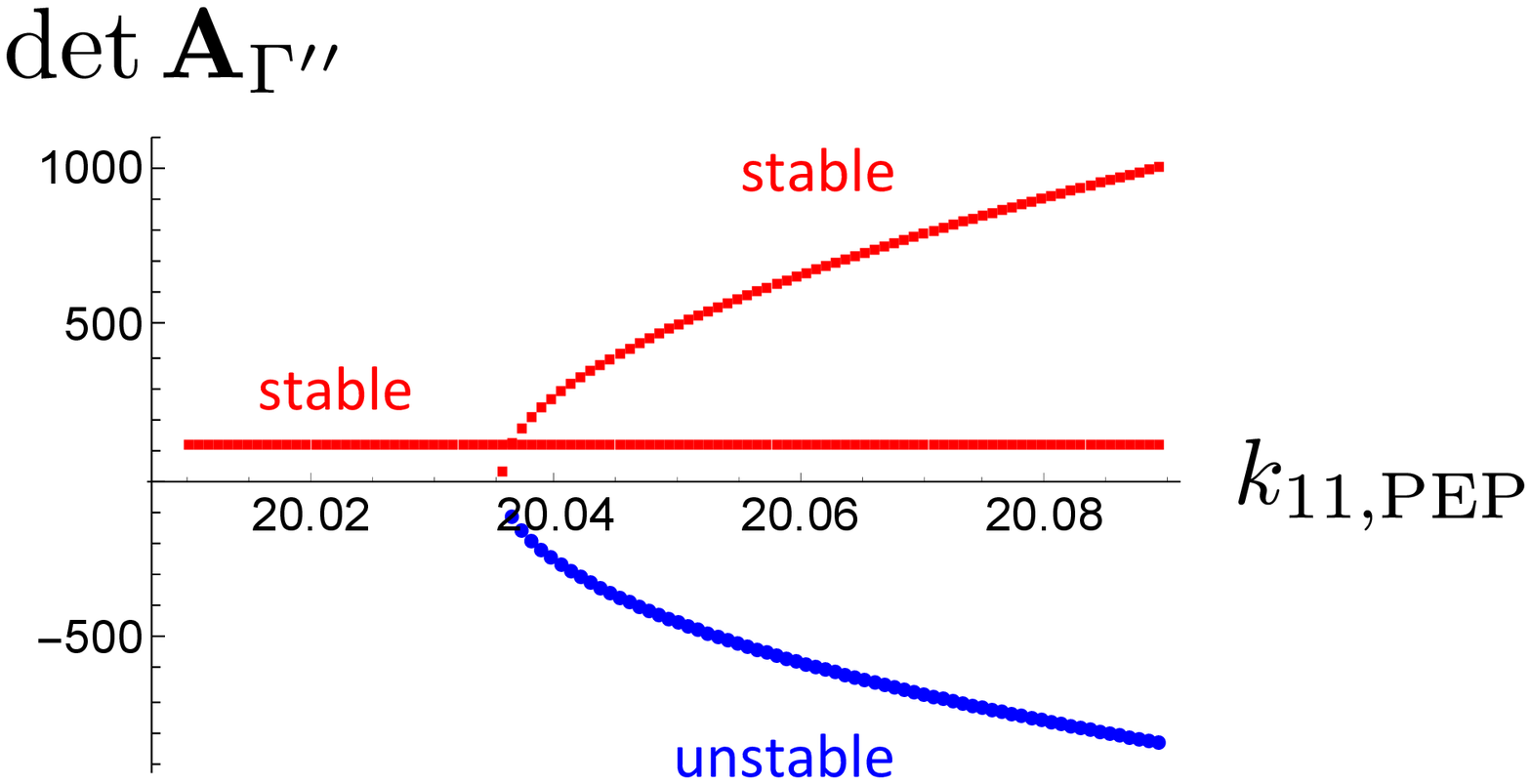}
\caption{ $\detA_{\Gamma''}$ versus $ k_{11,PEP}$. The  red curves correspond to two stable solutions, and the  blue curve corresponds to an unstable solution.   }
\label{fig:detAec}
\end{figure}

 Fig. \ref{fig:detAec} shows the numerical result for $\detA_{\Gamma''}$ versus  $ k_{11,PEP}$.  For large $ k_{11,PEP}$, there are two stable solutions (red curves) and one unstable solution (blue curve). As $ k_{11,PEP}$ is decreased,  the values of $\detA_{\Gamma''}$ for a stable and unstable solutions decreases and eventually approach zero. Thus, the parameter $ k_{11,PEP}$, which is one of the inducing parameters for $\Gamma''$,  actually induces a bifurcation associated with $\Gamma''$. 

\begin{figure}[h]
\includegraphics[width=14cm]{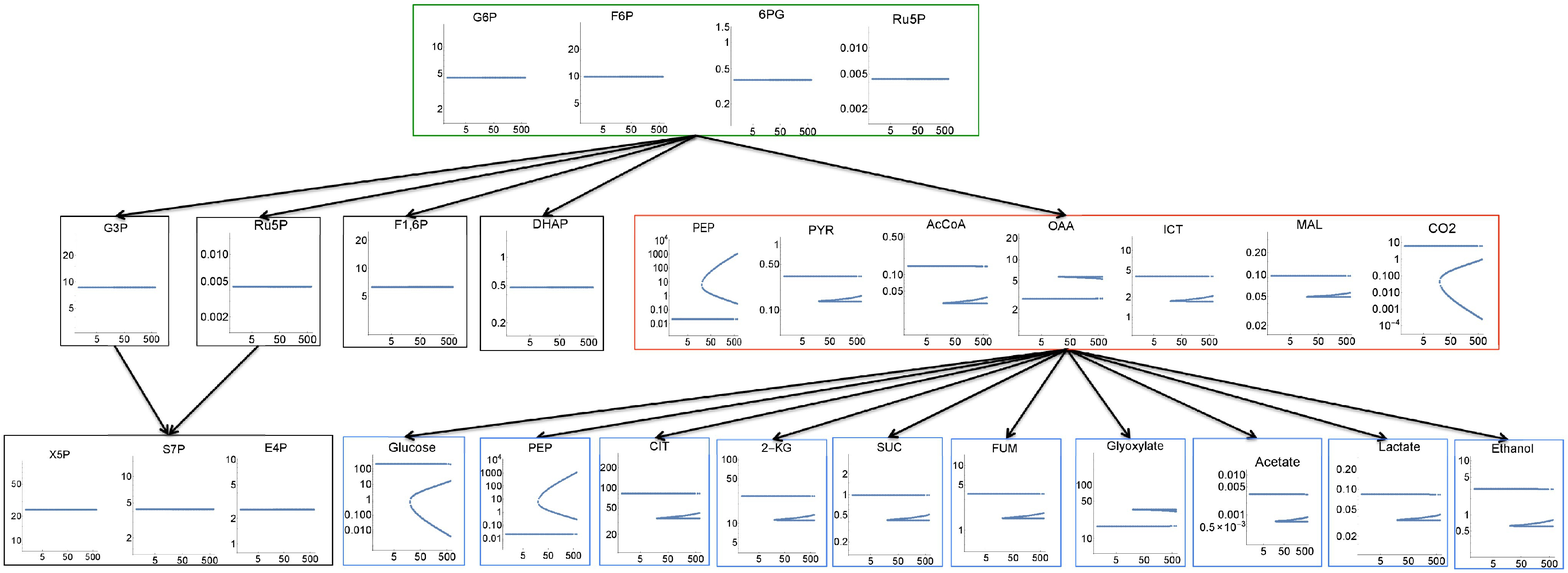}
\caption{ Steady-state concentrations versus $ k_{11,\rm PEP}$ (the horizontal axis) in the E. coli system. Each box corresponds to the one in Fig. \ref{fig:ecbs}.  A saddle-node bifurcation is observed for each of the bifurcating chemicals (the red and blue boxes) for  $\Gamma''$:  {Each chemical in the red and blue boxes has  one stable branch when  $ k_{11,\rm PEP}$ is small, and has two stable and one unstable branches when  $ k_{11,\rm PEP}$ is large.  Each chemical in the green and black boxes  is independent of  $ k_{11,\rm PEP}$, due to the law of localization, and has one stable branch. }}
\label{fig:ecplot}
\end{figure}

The bifurcating chemicals for $\Gamma''$ are those in the blue and red boxes.  Fig. \ref{fig:ecplot} shows the numerical results for the steady-state concentrations versus the parameter $ k_{11,\rm PEP}$. We see that  saddle-node bifurcations are observed only for the bifurcating chemicals. 

\section{An example exhibiting transcritical bifurcation}
\begin{figure}[t]
\center
\includegraphics[clip,width=4.5cm]{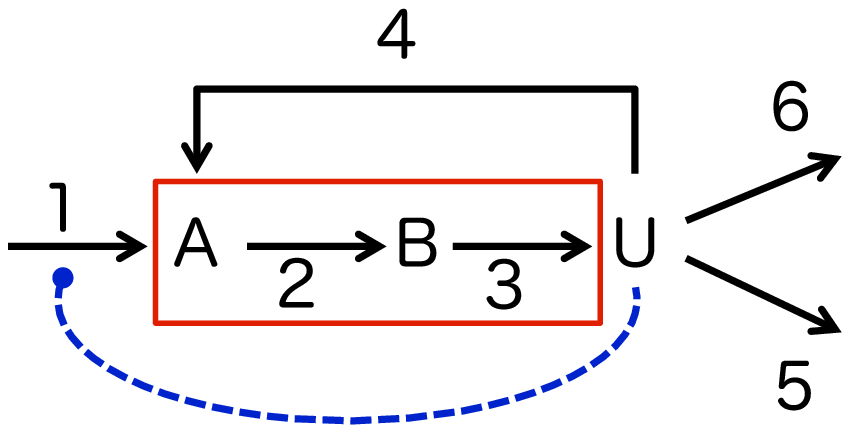}
\caption{System consisting of six reactions, ${\rm U}  \overset{k_1}{\rightarrow }  {\rm U+ A} , \ {\rm A}\overset{k_2}{\rightarrow } {\rm B},
{\rm B} \overset{k_3}{\rightarrow }{\rm U},\  {\rm U}\overset{k_4}{\rightarrow }{\rm A},
2 {\rm U} \overset{k_5}{\rightarrow}  {\rm U},  \ {\rm  U} \overset{k_6}{\rightarrow}$.
The red box indicates a buffering structure. }
\label{fig:tc}
\end{figure}

Consider the system in Fig. \ref{fig:tc}. This system is designed by modifying the First Schl\"ogl Model \cite{schlogl}.
%
It consists of reactions
\bal
{\rm U}  \overset{k_1}{\rightarrow }  {\rm U+ A} , \ {\rm A}\overset{k_2}{\rightarrow } {\rm B}, \ &
{\rm B} \overset{k_3}{\rightarrow }{\rm U},\  {\rm U}\overset{k_4}{\rightarrow }{\rm A},\non\\
2 {\rm U} \overset{k_5}{\rightarrow}  {\rm U}, & \ {\rm  U} \overset{k_6}{\rightarrow}.
\eal
The stoichiometry matrix $\nu$ is
\bal
\nu= \left(
\begin{array}{cccccc}
 1 & -1 & 0 & 1 & 0 & 0 \\
 0 & 1 & -1 & 0 & 0 & 0 \\
 0 & 0 & 1 & -1 & -1 & -1 \\
\end{array}
\right),
\eal
which has three independent kernel vectors ${\mbf c}_1= (1, 1, 1, 0, 0, 1)^T, {\mbf c}_2=(1, 1, 1, 0, 1, 0)^T, {\mbf c}_3=(0, 1, 1, 1, 0, 0)^T$.

The system  contains a buffering structure ${\Gamma_s}=(\{{\rm A,B}\},\{2,3 \})$, since $\chi = 2-2+0=0$.
By permutating the row index  as $\{2,3,1,4,5,6 \}$ and the column index  as $\{\text{A},\text{ B},\text{ U}, {\mbf c}_1,{\mbf c}_2,{\mbf c}_3\}$, we obtain
\bal
{ \bf A}
=\left(
\begin{array}{cccccc}
\cline{1-2}
\multicolumn{1}{|c}{ r_{2,\text{A}}} &\multicolumn{1}{c|}{ 0} & 0 & 1 & 1 & 1 \\
\multicolumn{1}{|c}{ 0} &\multicolumn{1}{c|}{  r_{3,\text{B}}} & 0 & 1 & 1 & 1 \\
\cline{1-6}
 0 & 0 & \multicolumn{1}{|c}{r_{1,\text{U}}} & 1 & 1 & \multicolumn{1}{c|}{ 0 }\\
 0 & 0 & \multicolumn{1}{|c}{r_{4,\text{U}} }& 0 & 0 & \multicolumn{1}{c|}{ 1} \\
 0 & 0 &\multicolumn{1}{|c}{ r_{5,\text{U}} }& 0 & 1 & \multicolumn{1}{c|}{ 0} \\
 0 & 0 & \multicolumn{1}{|c}{r_{6,\text{U}} }& 1 & 0 & \multicolumn{1}{c|}{ 0} \\
 \cline{3-6}
\end{array}
\right),
\eal
where the upper-left block and the lower-right block correspond to $\bf A_{{\Gamma_s}}$ and $\bf A_{\bar{\Gamma}_s}$, respectively.  Thus, the determinant becomes
\bal
\detA = \underset{\detA_{\Gamma_s}}{\underbrace{ r_{2,\text{A}} r_{3,\text{B}}}}  \underset{\detA_{\bar{\Gamma}_s}}{\underbrace{\left(-r_{1,\text{U}}+r_{5,\text{U}}+r_{6,\text{U}}\right)}}.
\eal
Thus, a steady-state bifurcation can occur  when $\detA_{\bar{\Gamma}_s}$  changes its sign.

\subsection{\red{The mass-action kinetics case}}
\begin{figure}[htbp]
\center
\includegraphics[clip,width=8cm]{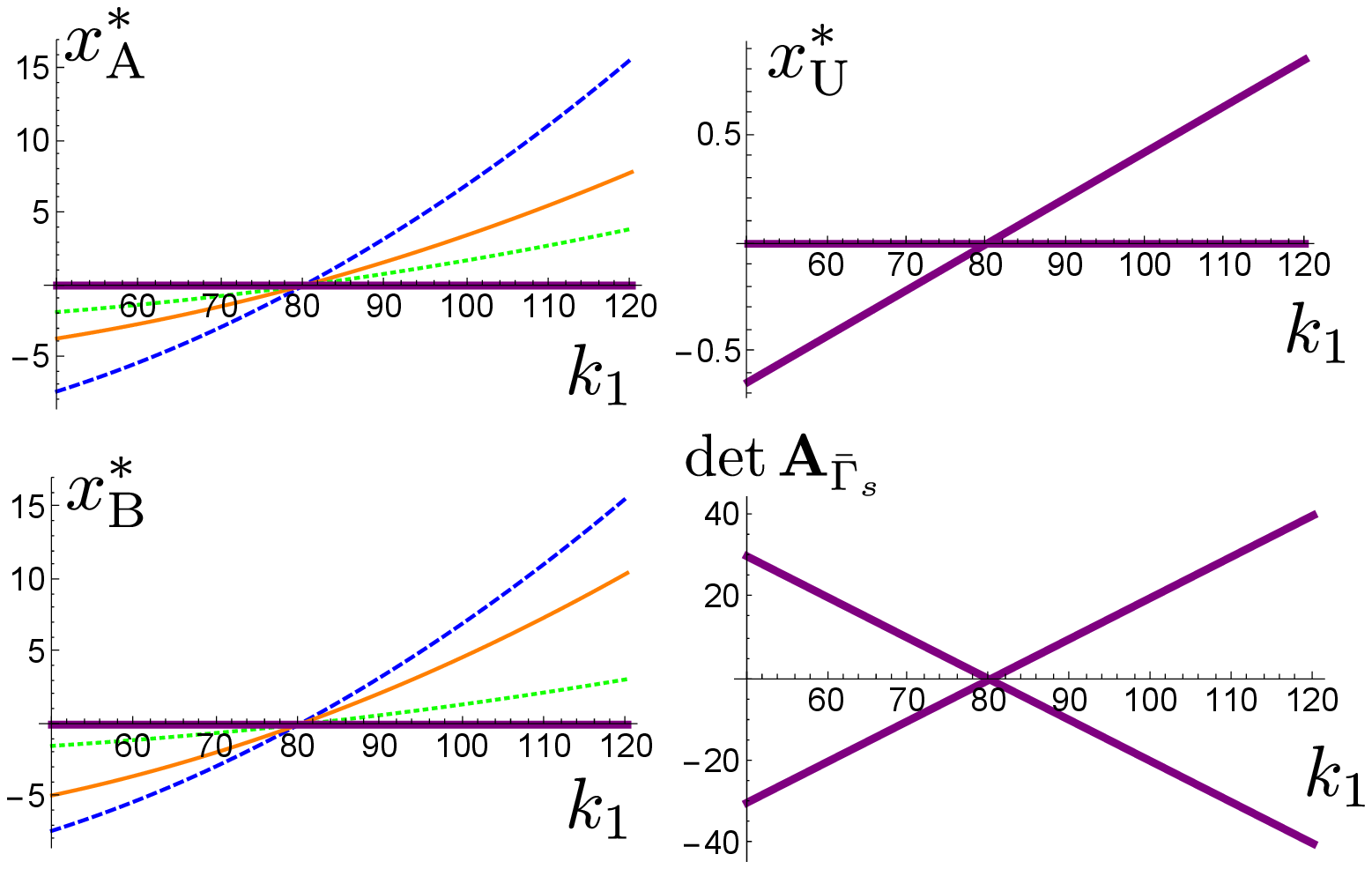}
\caption{Concentrations and $\detA_{\bar{\Gamma}_s}$ for different values of $(k_2,k_3)$.  
All  three cases have the solution ${\mbf x}^*={\mbf 0}$. The dashed blue line, the solid orange line, and the dotted green line correspond to $(k_2,k_3)=(10,10),\ (20,15),\ (40,50)$, while
$(k_4,k_5,k_6)$ are fixed as $(k_4,k_5,k_6)$$=(65, 47, 80)$ in all of the three cases. 
\corr{The solid purple lines indicate that the plots for  these three parameter choices coincide with each other.}
}
\label{fig:trans}
\end{figure}

\corr{The inducing parameters for $\gbs$ are those associated with reaction in $\gbs$, that is, $\{k_1,k_4, k_5,k_6 \}$.
Fig. \ref{fig:trans} shows the bifurcation diagram, where  the mass-action kinetics is assumed as in \cite{schlogl}. 
The determinant $\detA_\gbs$ and the steady-state concentrations versus the parameter $k_1$ are shown for three different  sets of parameters inside ${\Gamma_s}$, i.e. $k_2, k_3$.  From the plots for  $\detA_\gbs$,  we see  the parameter $k_1$, which is an inducing parameter for $\gbs$, actually induces the bifurcation associated with $\gbs$.}

\corr{ As for steady-state concentrations, the bifurcations are observed  in  all chemicals in the system, since the bifurcating chemicals for $\bar\Gamma_s$ are all chemicals. In the plots for concentrations in Fig. \ref{fig:trans},  we see that, for small $k_1$, the system has a single stable solution ${\bm x}^* = \bm 0$ and an (unphysical negative) unstable solution. For large $k_1$, the former becomes an unstable solution and the latter becomes a positive stable solution.  Thus, the system in Fig. \ref{fig:trans} exhibits the transcritical bifurcation.   We can also see that both steady-state value of $x_{\rm U}^*$ and the critical value of $k_1\in  \gbs$ are independent of parameters inside $\Gamma_s$.}

In  the case of the mass-action kinetic, we can  explicitly obtain the analytic solutions. 
\red{Indeed, the steady state is given by} 
\bal
 (x^*_A,x^*_B,x^*_U)=  (0,0,0),\quad
  \bigl(\frac{(k_1+k_4) (k_1-k_6)}{k_2 k_5}, \frac{(k_1+k_4) (k_1-k_6)}{k_3 k_5}, \frac{k_1-k_6}{k_5}\bigr),
\label{transsol}
\eal
\red{and the determinant of {\bf A} is given by}
\bal
\detA =\underset{\detA_{\Gamma_s}}{ \underbrace{k_2 k_3} }\underset{\detA_{\bar{\Gamma}_s}}{\underbrace{\left(-2 k_5 x^*_{\text{U}}+k_1-k_6\right) }} = \pm
     k_2 k_3 (k_1-k_6),
\eal
where $+$ and $-$ corresponds to the first and the second solution in \eqref{transsol}, respectively.
The critical point is given by $k_1 = k_6$, which is indeed independent \red{of   $k_2, k_3\in \gs$.}

\subsection{\red{The modified mass-action kinetics case}}
Finally, we  show the result for  a more complicated  kinetics.  Specifically, instead of the mass-action kinetics, we consider the following reaction rate functions,
\bal
(r_1,\ldots, r_6) = \left( k_1 x_{\text{U}},k_2  x_{\text{A}} \left(1+ \frac{k_{2,{\rm B}}x^2_{\text{B}}}{x^2_{\text{B}}+K }\right),k_3 x_{\text{B}} \left(1+ \frac{ k_{3,{\rm A} }x^2_{\text{A}}}{x^2_{\text{A}}+K}\right),k_4 x_{\text{U}},k_5 x_{\text{U}}^2,k_6 x_{\text{U}}\right)
\eal
This system reduces to the case of the mass-action kinetic  when $k_{2,{\rm B}}=0$ and $k_{3,{\rm A} }=0$.

%

\begin{figure}[htbp]
\centering
  \includegraphics[width=8.0cm]{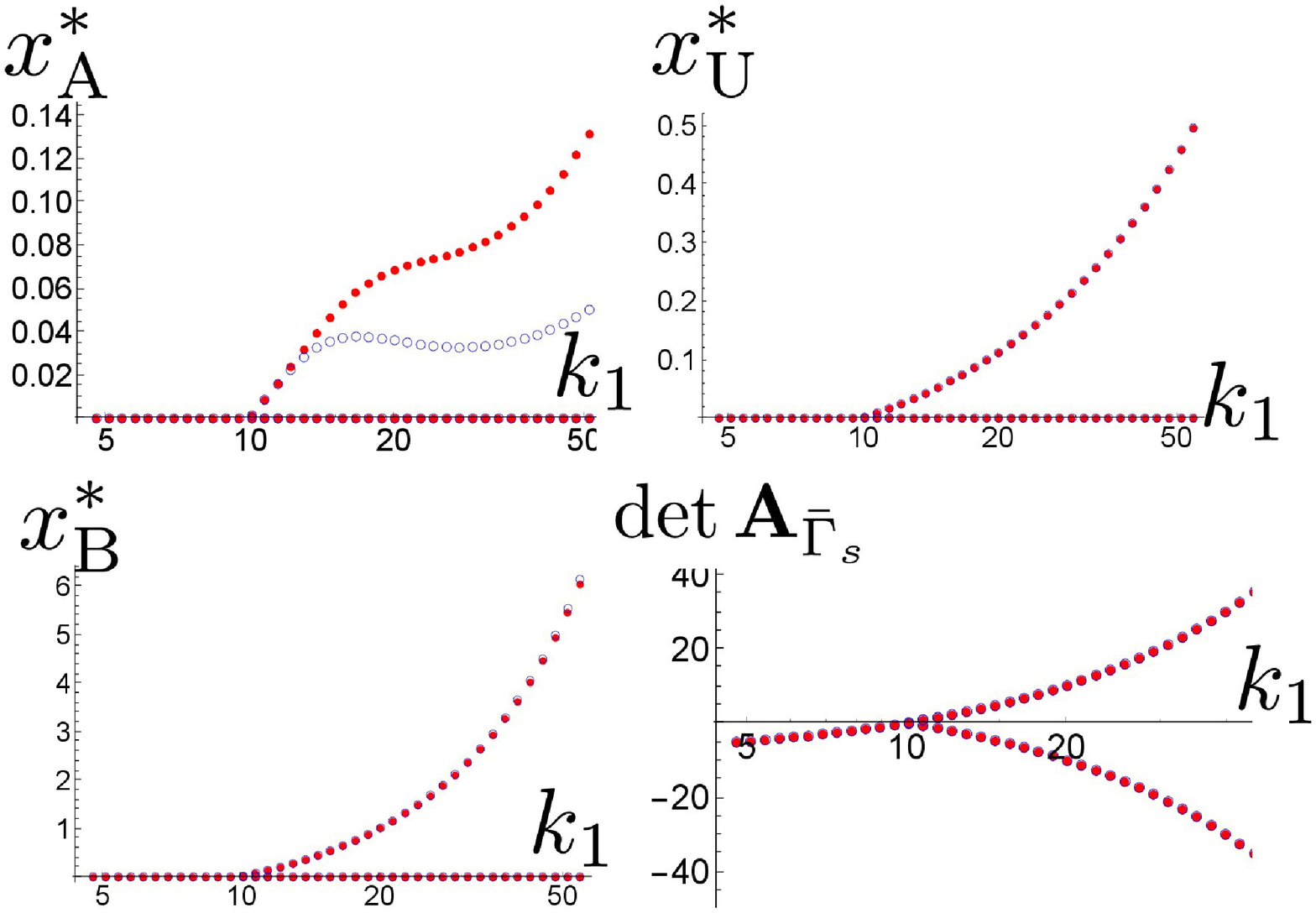}
   \caption{Concentrations and $\detA_{\gbs}$ for different values of $k_{2,B}$ and $k_{3,A}$. The solution ${\mbf x}^*=0$ exists for all parameters. The plots for A, B $\in \gs$  depend on  the parameters $k_{2,B},k_{3,A} \in \gs$, while those of ${\rm U}\in \gbs$  and $\detA_\gbs$ do not.   }
   \label{fig:tcapp}
\end{figure}

Fig. \ref{fig:tcapp} shows the numerical results of concentrations and $\detA$ for  different values of $k_{2,B}$ and $k_{3,A}$.
For the plot of \red{the curve of} filled circle,  $k_{2,B}=5$ and $k_{3,A}=5$.
For the plot of \red{the curve of} empty circle,  $k_{2,B}=10$ and $k_{3,A}=15$.
The other parameters are the same for the two plots; $(k_1,\ldots, k_6)=(40, 10, 80, 70, 90, 10)$ and $K=5$.
Note that only \red{non-negative physical} solutions are shown in Figure \ref{fig:tcapp}. We observe that the qualitative behavior is the same as  \red{the case of} the mass-action kinetics: $x^*_{{\rm U}}$, $\detA_{\bar \Gamma}$,
and the critical value of $k_1$ \red{is} independent of the parameters inside the buffering structure.  On the other hand,  the positive solutions of $x^*_{{\rm A}}$ and $x^*_{{\rm B}}$ depend on these parameters.

\section{Example network with $K_c>0 $}

The following system is known to exhibit a saddle-node bifurcation, when mass-action kinetics is assumed \cite{CC}.
 {\bal
\text{A} + \text{E1}& \underset{2}{ \overset{1}{\rightleftharpoons}} \text{ AE1}\non\\
\text{AE1} & \overset{3}{\rightarrow}  \text{Ap + E1}\non\\
\text{Ap + E1} &  \underset{5}{ \overset{4}{\rightleftharpoons}}  \text{ApE1}\non\\
\text{ApE1} &  \overset{6}{\rightarrow} \text{App + E1}\non\\
\text{App + E2} &   \underset{8}{ \overset{7}{\rightleftharpoons}}  \text{ AppE2}\non\\
\text{AppE2} &   \overset{9}{\rightarrow} \text{Ap + E2}\non\\
\text{Ap + E2}  &    \underset{11}{ \overset{10}{\rightleftharpoons}}    \text{ApE2}\non\\
\text{ApE2 }&   \overset{12}{\rightarrow}  \text{A + E2} \label{phosphos}
\eal
}

Here, we consider an  extended system  by coupling the above system \eqref{phosphos}  with the following four reactions,
\bal
\text{Ap  + B} & \underset{14}{ \overset{13}{\rightleftharpoons}} \text{A +  Bp} \non\\
\text{B + F} & \underset{16}{ \overset{15}{\rightleftharpoons}}  \text{BF }\label{ext}
\eal
In  this extended  system,  the extended part  $\Gamma_s = \text{(B,Bp,F,13,14,15,16)}$ is a buffering structure because
$\chi = 3 - 4 + 2 -1$.  The complement subnetwork $\gbs$ in the extended system consists of all the chemicals and reactions existing in the original system.

The stoichiometry matrix is given by
\bal
\nu = \left(
\begin{array}{cccccccccccccccc}
 -1 & 1 & 0 & 0 & 0 & 0 & 0 & 0 & 0 & 0 & 0 & 1 & 1 & -1 & 0 & 0 \\
 -1 & 1 & 1 & -1 & 1 & 1 & 0 & 0 & 0 & 0 & 0 & 0 & 0 & 0 & 0 & 0 \\
 1 & -1 & -1 & 0 & 0 & 0 & 0 & 0 & 0 & 0 & 0 & 0 & 0 & 0 & 0 & 0 \\
 0 & 0 & 1 & -1 & 1 & 0 & 0 & 0 & 1 & -1 & 1 & 0 & -1 & 1 & 0 & 0 \\
 0 & 0 & 0 & 1 & -1 & -1 & 0 & 0 & 0 & 0 & 0 & 0 & 0 & 0 & 0 & 0 \\
 0 & 0 & 0 & 0 & 0 & 1 & -1 & 1 & 0 & 0 & 0 & 0 & 0 & 0 & 0 & 0 \\
 0 & 0 & 0 & 0 & 0 & 0 & -1 & 1 & 1 & -1 & 1 & 1 & 0 & 0 & 0 & 0 \\
 0 & 0 & 0 & 0 & 0 & 0 & 1 & -1 & -1 & 0 & 0 & 0 & 0 & 0 & 0 & 0 \\
 0 & 0 & 0 & 0 & 0 & 0 & 0 & 0 & 0 & 1 & -1 & -1 & 0 & 0 & 0 & 0 \\
 0 & 0 & 0 & 0 & 0 & 0 & 0 & 0 & 0 & 0 & 0 & 0 & -1 & 1 & -1 & 1 \\
 0 & 0 & 0 & 0 & 0 & 0 & 0 & 0 & 0 & 0 & 0 & 0 & 1 & -1 & 0 & 0 \\
 0 & 0 & 0 & 0 & 0 & 0 & 0 & 0 & 0 & 0 & 0 & 0 & 0 & 0 & -1 & 1 \\
 0 & 0 & 0 & 0 & 0 & 0 & 0 & 0 & 0 & 0 & 0 & 0 & 0 & 0 & 1 & -1 \\
\end{array}
\right),
\eal
where  the column indices from left to right are
\bal
\{\text{A},\text{E1},\text{AE1},\text{Ap},\text{ApE1},\text{App},\text{E2},\text{AppE2},\text{ApE2},\text{B},\text{Bp},\text{F},\text{BF}\}.
\eal
We represent the rate constants for the twelve reactions in \red{\eqref{phosphos}} as $k_1,\ldots k_{12}$, 
and the four reactions in \eqref{ext} as $k_{13},\ldots k_{16}$. 
For example, the rate functions of the first and the second reaction are $R_1 = k_1 x_{A} x_{E1}$ \red{and} $R_2 = k_2 x_{AE1}$, respectively.

The  matrix $\bf A$ is  \red{given by}
{\scriptsize
\bal
\left(
\begin{array}{ccccccccccccc|cccccccc}
k_1 x_{\text{E1}} & k_1 x_{\text{A}} & 0 & 0 & 0 & 0 & 0 & 0 & 0 & 0 & 0 & 0 & 0 & 0 & 0 & 1 & 0 & 0 & 0 & 0 & 1 \\
 0 & 0 & k_2 & 0 & 0 & 0 & 0 & 0 & 0 & 0 & 0 & 0 & 0 & 0 & 0 & 0 & 0 & 0 & 0 & 0 & 1 \\
 0 & 0 & k_3 & 0 & 0 & 0 & 0 & 0 & 0 & 0 & 0 & 0 & 0 & 0 & 0 & 1 & 0 & 0 & 0 & 0 & 0 \\
 0 & k_4 x_{\text{Ap}} & 0 & k_4 x_{\text{E1}} & 0 & 0 & 0 & 0 & 0 & 0 & 0 & 0 & 0 & 0 & 0 & 0 & 0 & 1 & 0 & 1 & 0 \\
 0 & 0 & 0 & 0 & k_5 & 0 & 0 & 0 & 0 & 0 & 0 & 0 & 0 & 0 & 0 & 0 & 0 & 0 & 0 & 1 & 0 \\
 0 & 0 & 0 & 0 & k_6 & 0 & 0 & 0 & 0 & 0 & 0 & 0 & 0 & 0 & 0 & 0 & 0 & 1 & 0 & 0 & 0 \\
 0 & 0 & 0 & 0 & 0 & k_7 x_{\text{E2}} & k_7 x_{\text{App}} & 0 & 0 & 0 & 0 & 0 & 0 & 0 & 0 & 0 & 0 & 1 & 1 & 0 & 0 \\
 0 & 0 & 0 & 0 & 0 & 0 & 0 & k_8 & 0 & 0 & 0 & 0 & 0 & 0 & 0 & 0 & 0 & 0 & 1 & 0 & 0 \\
 0 & 0 & 0 & 0 & 0 & 0 & 0 & k_9 & 0 & 0 & 0 & 0 & 0 & 0 & 0 & 0 & 0 & 1 & 0 & 0 & 0 \\
 0 & 0 & 0 & k_{10} x_{\text{E2}} & 0 & 0 & k_{10} x_{\text{Ap}} & 0 & 0 & 0 & 0 & 0 & 0 & 0 & 0 & 1 & 1 & 0 & 0 & 0 & 0 \\
 0 & 0 & 0 & 0 & 0 & 0 & 0 & 0 & k_{11} & 0 & 0 & 0 & 0 & 0 & 0 & 0 & 1 & 0 & 0 & 0 & 0 \\
 0 & 0 & 0 & 0 & 0 & 0 & 0 & 0 & k_{12} & 0 & 0 & 0 & 0 & 0 & 0 & 1 & 0 & 0 & 0 & 0 & 0 \\
 0 & 0 & 0 & k_{13} x_{\text{B}} & 0 & 0 & 0 & 0 & 0 & k_{13} x_{\text{Ap}} & 0 & 0 & 0 & 0 & 1 & 0 & 0 & 0 & 0 & 0 & 0 \\
 k_{14} x_{\text{Bp}} & 0 & 0 & 0 & 0 & 0 & 0 & 0 & 0 & 0 & k_{14} x_{\text{A}} & 0 & 0 & 0 & 1 & 0 & 0 & 0 & 0 & 0 & 0 \\
 0 & 0 & 0 & 0 & 0 & 0 & 0 & 0 & 0 & k_{15} x_{\text{F1}} & 0 & k_{15} x_{\text{B}} & 0 & 1 & 0 & 0 & 0 & 0 & 0 & 0 & 0 \\
 0 & 0 & 0 & 0 & 0 & 0 & 0 & 0 & 0 & 0 & 0 & 0 & k_{16} & 1 & 0 & 0 & 0 & 0 & 0 & 0 & 0 \\ \\ \hline
  1 & 0 & 1 & 1 & 1 & 1 & 0 & 1 & 1 & 0 & 0 & 0 & 0 & 0 & 0 & 0 & 0 & 0 & 0 & 0 & 0 \\
 0 & 1 & 1 & 0 & 1 & 0 & 0 & 0 & 0 & 0 & 0 & 0 & 0 & 0 & 0 & 0 & 0 & 0 & 0 & 0 & 0 \\
 0 & 0 & 0 & 0 & 0 & 0 & 1 & 1 & 1 & 0 & 0 & 0 & 0 & 0 & 0 & 0 & 0 & 0 & 0 & 0 & 0 \\
 0 & 0 & 0 & 0 & 0 & 0 & 0 & 0 & 0 & 1 & 1 & 0 & 1 & 0 & 0 & 0 & 0 & 0 & 0 & 0 & 0 \\
 0 & 0 & 0 & 0 & 0 & 0 & 0 & 0 & 0 & 0 & 0 & 1 & 1 & 0 & 0 & 0 & 0 & 0 & 0 & 0 & 0 \\
\end{array}
\right),
\eal
}
and the \red{corresponding determinant $\detA$ is} {\scriptsize
\bal
\detA= &\underset{{\rm det}\, {\bf A}_\gs }{\underbrace{ \biggl(k_{14} x_{\text{A}} (k_{15} (x_{\text{B}}+x_{\text{F1}})+k_{16})+k_{13} x_{\text{Ap}} (k_{15} x_{\text{B}}+k_{16})\biggr)}} \\
 &\times \biggl(k_1 (k_3 x_{\text{E1}} (k_4 (k_7 k_9 x_{\text{E2}} (k_{10} x_{\text{Ap}} (x_{\text{A}}+x_{\text{Ap}}+x_{\text{App}}+x_{\text{E1}}\non\\
 &+x_{\text{E2}})+k_{11} (x_{\text{A}}+x_{\text{Ap}}+x_{\text{E1}})+k_{12} (x_{\text{A}}+x_{\text{Ap}}+x_{\text{E1}}))\non\\
 &+k_6 x_{\text{E1}} (k_8 (k_{10} x_{\text{Ap}}+k_{11}+k_{12})+k_9 (k_{10} x_{\text{Ap}}+k_{11}+k_{12})+k_7 (k_{11}+k_{12})\non\\
 & (x_{\text{App}}+x_{\text{E2}})))+(k_5+k_6) k_7 k_9 x_{\text{E2}} (k_{10} (x_{\text{Ap}}+x_{\text{App}}+x_{\text{E2}})+k_{11}+k_{12}))\non\\
& -k_{10} k_{12} x_{\text{E2}} (k_4 k_6 x_{\text{Ap}} x_{\text{E1}} (k_7 (x_{\text{A}}+x_{\text{Ap}}+x_{\text{App}}+x_{\text{E1}}\non\\
 &+x_{\text{E2}})+k_8+k_9)-(k_5+k_6) k_7 k_9 x_{\text{E2}} (x_{\text{A}}+x_{\text{E1}})))-(k_2+k_3) k_7 k_{10} k_{12} x_{\text{E2}}\non\\
 & (k_4 k_6 x_{\text{Ap}} x_{\text{E1}}-k_9 x_{\text{E2}} (k_4 x_{\text{Ap}}+k_5+k_6))\biggr).
\eal
}

While the factor ${\rm det}\, {\bf A}_\gs$ is positive definite,  the factor ${\rm det}\,{\bf A}_{\gbs}$ can change the sign as we change parameters since it contains the negative terms. Therefore, the saddle-node bifurcations can arise from $\detA_\gbs$. Furthermore, the factor ${\rm det}\, {\bf A}_\gbs$ is  independent of  the parameters and concentrations in $\Gamma$, as is ensured from the property of  buffering structure. Therefore, the bifurcation points of this extended system are completely the same as the original system \eqref{phosphos}. Below, we demonstrate these expectations by numerical computation.

\begin{figure}[htbp]
\centering
  \includegraphics[width=10.0cm]{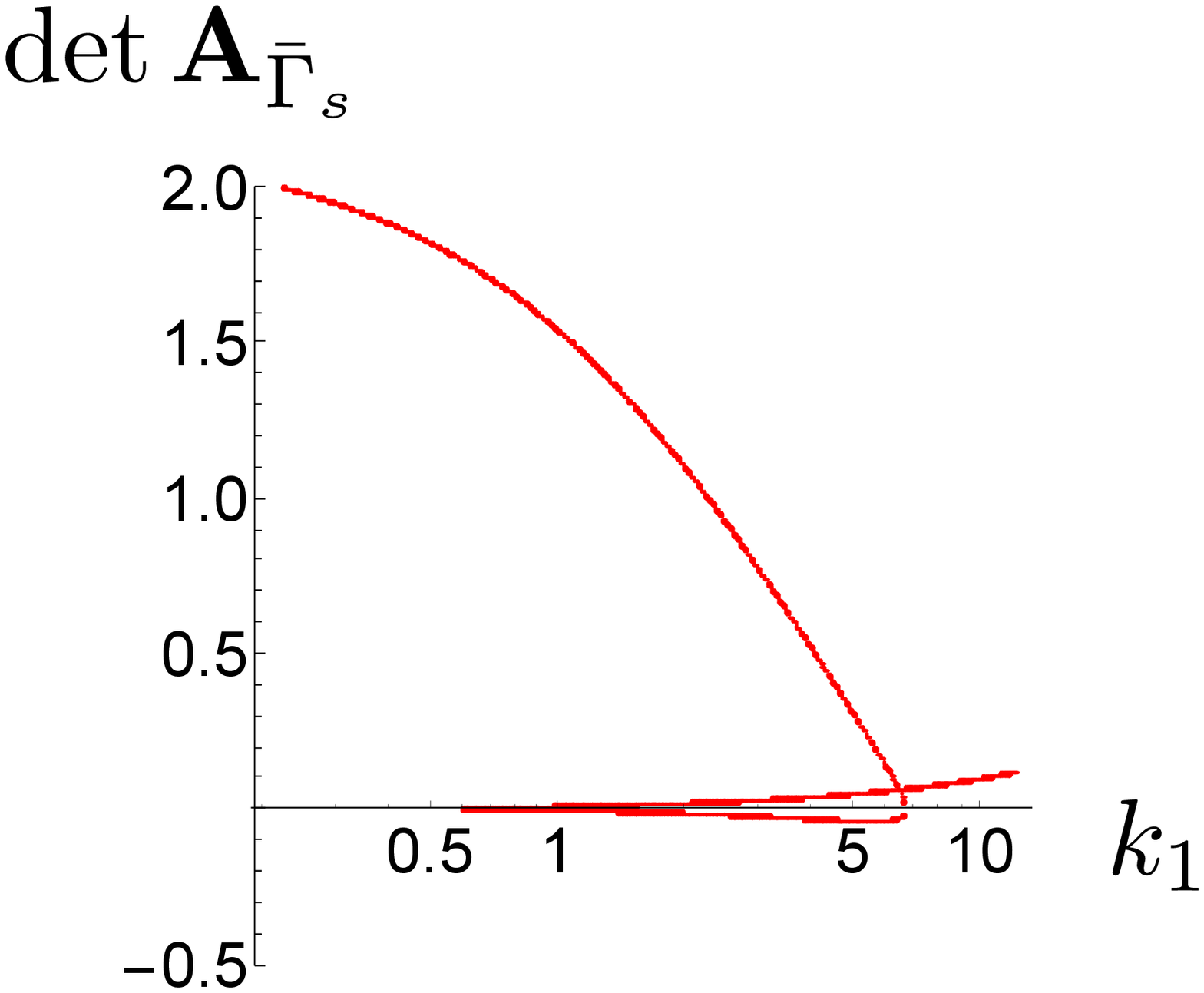}
   \caption{Values of  ${\rm det} \bf A_\gbs$ versus $k_1$. }
   \label{fig:detA}
\end{figure}

\corr{The inducing parameters for the subnetwork $\gbs$ are parameters in $\gbs$. Fig. \ref{fig:detA} shows that the determinant of $\bf A$ versus the parameter $k_1$, which is the parameter in $\gbs$.  The other  rata parameters are chosen as
\bal
(k_2,\ldots, k_{12}) &= (30, 30, 20, 2, 2, 800, 1, 1, 500, 14, 14), \non\\
(k_{13}, \ldots, k_{16})& =(3, 2, 10, 7)
\eal
and the  values of conserved concentrations are chosen as
\bal
x_{\text{A}}+x_{\text{AE1}}+x_{\text{Ap}}+x_{\text{ApE1}}+x_{\text{ApE2}}+x_{\text{App}}+x_{\text{AppE2}}=10\non\\
x_{\text{AE1}}+x_{\text{ApE1}}+x_{\text{E1}}=3\non\\
x_{\text{ApE2}}+x_{\text{AppE2}}+x_{\text{E2}}=4\non\\
x_{\text{B}}+x_{\text{BF1}}+x_{\text{Bp}}=6\non\\
x_{\text{BF1}}+x_{\text{F1}}=4.
\eal
In Fig. \ref{fig:detA}, we see that, as we change the parameter $k_1$,   there are two bifurcation points with $\detA_\gbs=0$. Thus, the parameter $k_1$ indeed acts as a bifurcation parameter associated with the subnetwork $\gbs$. }

\begin{figure}[h]
\centering
   \includegraphics[width=10.0cm]{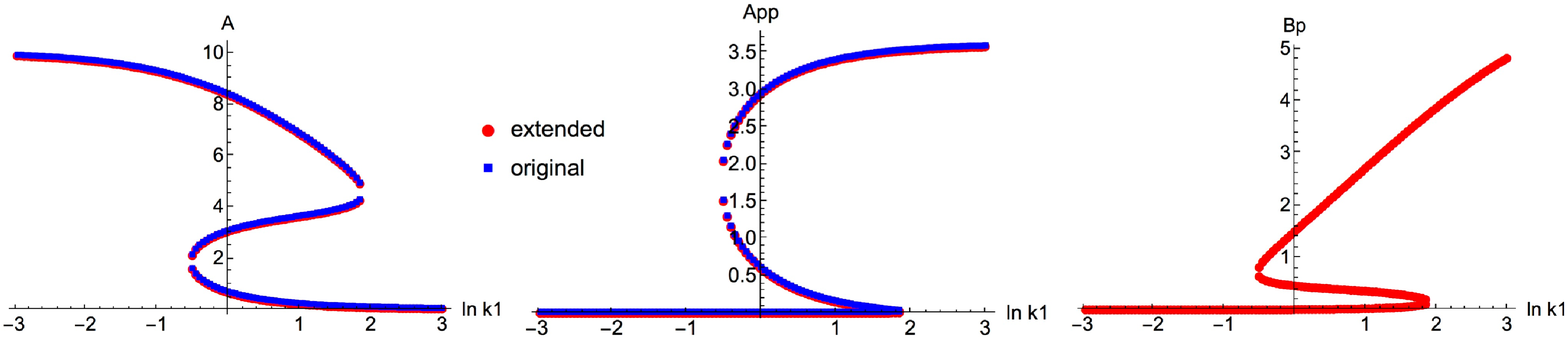}
   \caption{(Left,Center) Steady-state concentrations of $A, App$ for the original and the extended system. (Right) Concentration of $Bp$ in the extended system. }
   \label{fig:phosphos}
\end{figure}

\corr{The bifurcation chemicals for $\gbs$ are all chemicals in the extended system. This is illustrated in 
Fig. \ref{fig:phosphos}, where  the steady-state concentrations for $A, App$ and  $Bp$ versus $k_1$ are shown. 
We also see that the original and extended systems have exactly the same critical value of $k_1$, as expected. Furthermore, for chemicals  existing in the original system \eqref{phosphos}, or equivalently chemicals in $\gbs$,  the plots for the extended system   exactly  coincides with  those in the original system. This can be understood from the law of localization:  The original system can be obtained from the extended system,  by taking  the limit that the parameters in $\gs$ (i.e. $k_{13},k_{14},k_{15},k_{16}$)  go to zero.  However, the chemicals in $\gbs$  are not influenced in the limiting procedure because the law of localization states that  changing the parameters in $\gs$ does not influence the chemicals in $\gbs$. }

\bibliographystyle{aip}

\end{document}